\documentclass[3p,a4paper, 12pt,sort&compress]{elsarticle}

\usepackage{slashed}
\usepackage{bm}
\usepackage{braket}
\usepackage{amsmath,amssymb,amsfonts}
\usepackage{color}

\usepackage{textpos}

\newcommand{\D}{\mathrm{d}}

\begin{document}

\begin{frontmatter}
\title{{\bfseries Constraining the effective action by a \\ method of external sources}}
\author[a]{Bj\"{o}rn Garbrecht}
\ead{garbrecht@tum.de}
\author[a,b]{Peter Millington}
\ead{p.millington@nottingham.ac.uk}
\address[a]{Physik-Department T70, James-Franck-Stra\ss e,\\
Technische Universit\"{a}t M\"{u}nchen, 85748 Garching, Germany}
\address[b]{School of Physics and Astronomy, University of Nottingham,\\
Nottingham NG7 2RD, United Kingdom}

\begin{abstract}
We propose a novel method of evaluating the effective action, wherein the physical one- and two-point functions are obtained in the limit of non-vanishing external sources. We illustrate the self-consistency of this method by recovering the usual 2PI effective action due to Cornwall, Jackiw and Tomboulis, differing only by the fact that the saddle-point evaluation of the path integral is performed along the extremal \emph{quantum}, rather than \emph{classical}, path. As such, this approach is of particular relevance to situations where the dominant quantum and classical paths are non-perturbatively far away from one-another. A pertinent example is the decay of false vacua in radiatively-generated potentials, as may occur for the electroweak vacuum of the Standard Model. In addition, we describe how the external sources may instead be chosen so as to yield the two-particle-point-irreducible (2PPI) effective action of Coppens and Verschelde. Finally, in the spirit of the symmetry-improved effective action of Pilaftsis and Teresi, we give an example of how the external sources can be used to preserve global symmetries in truncations of the 2PI effective action. Specifically, in the context of an $O(2)$ model with spontaneous symmetry breaking, we show that this approach allows the Hartree-Fock approximation to be re-organized, such that the Goldstone boson remains massless algebraically in the symmetry-broken phase and we obtain the correct second-order thermal phase transition.
\end{abstract}



\end{frontmatter}

\begin{textblock}{14}(3,-20.5)
\begin{flushright}
TUM-HEP-1016-15
\end{flushright}
\end{textblock}

\section{Introduction}

The use of effective-action techniques has become ubiquitous across theoretical physics, both in the relativistic regime of high-energy processes and the non-relativistic setting of condensed matter systems. Such techniques play a significant role in the study of both perturbative and non-perturbative effects, including phase transitions, transport phenomena and renormalization group evolution.

The functional evaluation of the one-particle-irreducible (1PI) effective action was first described by Jackiw~\cite{Jackiw:1974cv} and subsequently generalized to nPI by Cornwall, Jackiw and Tomboulis (CJT)~\cite{Cornwall:1974vz}. The effective action provides a systematic means for obtaining the quantum equations of motion for $n$-point correlation functions, which automatically resum infinite sets of diagrams. However, in order to make the solution of these systems of equations tractable, we must, in reality, find consistent truncation schemes that preserve the underlying symmetries, and much attention has been given to this in the literature.

It is well known that truncations of the 2PI effective action do not, in general, preserve global and local symmetries of the effective action~\cite{Arrizabalaga:2002hn,Carrington:2003ut} due to higher-order effects. The reason for this can be understood heuristically as follows~\cite{Mottola:2003vx}: the satisfaction of symmetry identities, such as the Ward-Takahashi identities~\cite{Ward:1950xp,Takahashi:1957xn}, in the case of global and Abelian gauge theories, or the Slavnov-Taylor identities~\cite{Taylor:1971ff,Slavnov:1972fg}, in the case of non-Abelian gauge theories, requires the cancellation of diagrams of different topologies. However, once the 2PI effective action has been truncated at some finite order in the loop expansion, only a subset of all topologies are resummed, and the required cancellation is no longer exact. In the case of global symmetries, such as $O(N)$ models with spontaneous symmetry breaking (SSB), this problem manifests in the violation of Goldstone's theorem~\cite{Goldstone:1961eq,Goldstone:1962es}, with the Goldstone bosons acquiring non-zero masses in the SSB phase~\cite{Baym:1977qb,AmelinoCamelia:1997dd,Petropoulos:1998gt,Lenaghan:1999si}. A number of authors have proposed solutions to this problem~\cite{Petropoulos:1998gt,Lenaghan:1999si,Baacke:2002pi,Ivanov:2005yj, Ivanov:2005bv,Seel:2011ju,Grahl:2011yk,Marko:2013lxa}. These include the so-called external-propagator method~\cite{Nemoto:1999qf,vanHees:2002bv}, Optimized Perturbation Theory (OPT)~(see e.g.~refs.~\cite{Chiku:1998kd,Duarte:2011ph}) and the symmetry-improved CJT effective action~\cite{Pilaftsis:2013xna} of Pilaftsis and Teresi (PT), in which the Ward identities are imposed through the method of Lagrange multipliers. The latter variant of the effective action has the advantage that, aside from ensuring the masslessness of the Goldstone boson in the SSB phase and the correct second-order phase transition~\cite{Tetradis:1992xd}, it also yields the correct decay thresholds for both the Higgs and Goldstone modes. In addition, this approach has been shown to be free of the problem of IR divergences~\cite{Pilaftsis:2015cka,Pilaftsis:2015bbs} that arise as a result of the massless Goldstone bosons. In the case of QED, truncation of the 2PI effective action leads to violation of the Ward-Takahashi identities, and the transversality of the photon polarization cannot be guaranteed~\cite{Reinosa:2007vi}.

Once embedded in the Schwinger-Keldysh closed-time-path (CTP) formalism~\cite{Schwinger:1961,Keldysh:1964} of non-equilibrium field theory (see also refs.~\cite{Norton:1974bm,Jordan:1986ug,Calzetta:1986ey, Calzetta:1986cq,AmelinoCamelia:1992nc}), the CJT effective action allows the derivation of quantum transport equations by means of the Kadanoff-Baym formalism~\cite{Baym:1961zz, KB} (see also refs.~\cite{Blaizot:2001nr, Prokopec:2003pj, Prokopec:2004ic, Berges:2004yj, Millington:2012pf}). In recent years, these approaches have received a wealth of interest, not least in applications to the evolution of number densities in the early Universe. In addition, substantial progress has been made in the non-perturbative renormalization of the effective action both at zero and finite temperature~\cite{vanHees:2001ik,VanHees:2001pf,vanHees:2002bv,Blaizot:2003br, Blaizot:2003an,Berges:2004hn,Berges:2005hc}. In such cases, the physical limit of the effective action is obtained with non-vanishing external sources, where these encode information about the statistical ensemble of the system (see e.g.~ref.~\cite{Calzetta:1986cq} and also refs.~\cite{Berges:2004yj,Millington:2012pf}).

The discovery of a $\sim 125.5$ GeV Higgs boson~\cite{Aad:2012tfa, Chatrchyan:2012ufa,Agashe:2014kda} has brought into question the stability of the electroweak vacuum of the Standard Model~\cite{Cabibbo:1979ay, Sher:1988mj, Sher:1993mf,
Isidori:2001bm}. At present, state-of-the-art calculations~\cite{EliasMiro:2011aa,Degrassi:2012ry,Alekhin:2012py, Buttazzo:2013uya,Bednyakov:2015sca} (for a recent overview, see ref.~\cite{DiLuzio:2015iua}) suggest that the electroweak vacuum is metastable, having a lifetime longer than the current age of the Universe. However, these estimates are subject to a number of uncertainties. These include the current precision on the experimental determination of the top-quark mass~\cite{Bezrukov:2012sa,Masina:2012tz} and the potential impact of new physics at high scales~\cite{Branchina:2013jra, Branchina:2014usa, Branchina:2014rva, Lalak, Eichhorn:2015kea,Branchina:2015nda}. In addition, the standard RG improvement of the effective potential has recently been challenged~\cite{Gies:2014xha}. Finally, it remains an open question as to the correct method by which to determine the tunneling rate when the global minimum of the potential is generated radiatively~\cite{Weinberg:1992ds}. This latter issue is relevant also to the Coleman-Weinberg scenario of SSB~\cite{Coleman:1973jx}, as well as symmetry restoration at finite temperature~\cite{Kirzhnits:1972ut,
Dolan:1973qd,  Weinberg:1974hy}. In these cases, the extremal quantum path of the system is non-perturbatively far away from the extremal classical path,\footnote{In view of potential applications to vacuum metastability, we imply here that the most probable exit paths that are quantum in Minkowski space can be treated as classical in Euclidean space.} and it is therefore necessary to find ways of accounting consistently for the impact of these non-perturbative effects on the structure of the path integral itself. It is this observation that was the original motivation for this work.

The content of this article is as follows. In section~\ref{sec:ext}, we outline the method of evaluating the effective action in which the physical limit is obtained in the presence of \emph{non-vanishing} external sources in vacuum. This is in stark contrast to the standard approach, where the physical limit instead corresponds to \emph{vanishing} external sources. The external sources may then be used to constrain the effective action, and we will consider three concrete examples:
\begin{enumerate}
\item[(i)] In section~\ref{sec:CJT}, we choose the external sources such that the theory is forced along the extremal quantum path, recovering the 2PI CJT effective action~\cite{Cornwall:1974vz}. Subsequently, we discuss the relevance of this method to the problem of vacuum decay in potentials where the global minimum emerges only as a result of radiative corrections.

\item[(ii)] In section~\ref{sec:VC}, we show how the external sources may be chosen so as to recover the two-particle-point-irreducible (2PPI) effective action of Coppens and Verschelde (CV)~\cite{Verschelde:1992bs, Verschelde:1992ig, Coppens:1993zc, Coppens:1993ri}, which resums only local self-energy insertions. In addition, we will see that this method of external sources will allow us to avoid the problem of double-counting diagrams in the resummation without the need to isolate terms in the effective action artificially.

\item [(iii)] In section~\ref{sec:syms}, we consider the Hartree-Fock approximation~\cite{Hartree,Fock} in the case of a globally $O(2)$-invariant model with SSB. Therein, we obtain results in the spirit of the PT symmetry-improved effective action~\cite{Pilaftsis:2013xna,Pilaftsis:2015cka,Pilaftsis:2015bbs} and show that the Ward identity may be used to constrain the external sources so as to ensure that the Goldstone boson remains massless algebraically in the SSB phase. In addition, we will also show that this approach yields the correct second-order thermal phase transition, making a strong case for the utility of this novel method.
\end{enumerate}
Finally, our conclusions and potential future directions are presented in section~\ref{sec:conc}.

\section{Method of external sources}
\label{sec:ext}

In this section, we describe an alternative method of evaluating the effective action in which the physical limit is obtained for \emph{non-vanishing} external sources. We will make comparison with the standard approach, where the physical limit is instead obtained for \emph{vanishing} external sources. For the purposes of illustration, it will suffice to truncate the effective action at the level of the bi-local source. Nevertheless, this approach generalizes straightforwardly to the inclusion of higher poly-local sources. Furthermore, in order to present the diagrammatic structures explicitly, we will work at the two-loop level, truncating the effective action at order $\hbar^2$. The expansion at higher-loop orders will proceed analogously as per the general arguments that apply to the 2PI and 2PPI effective actions, respectively.

Throughout this article, we will work in four-dimensional Euclidean spacetime. The results derived, however, apply equally in Minkowski spacetime. Finally, for definiteness, we consider the $\lambda\Phi^4$ theory, with the Euclidean Lagrangian density
\begin{equation}
\label{eq:Lag}
\mathcal{L}_x\ =\ \frac{1}{2!}\big(\partial_{\mu}\,\Phi_x\big)^2\:+\:\frac{1}{2!}\,m^2\,\Phi_x^2\:+\:\frac{1}{4!}\,\lambda\,\Phi_x^4\;,
\end{equation}
where $\Phi_x\equiv \Phi(x)$ is a real scalar field of mass $m$, $\partial_{\mu}\equiv\partial/\partial x_{\mu}$ denotes the partial derivative with respect to the four-dimensional Euclidean spacetime coordinate $x_{\mu}=(\mathbf{x},\:x_4)$, and $\lambda$ is a dimensionless coupling.

In the presence of both a local and a bi-local source, the effective action $\Gamma[\phi,\Delta]$ is defined as the Legendre transform~\cite{Cornwall:1974vz}
\begin{equation}
\label{eq:Leg}
\Gamma[\phi,\Delta] \ \equiv\ \mathrm{max}_{J,K}\,\Gamma_{J,K}[\phi,\Delta]\ =\ \mathrm{max}_{J,K}\bigg[-\,\hbar\,\mathrm{ln}\,Z[J,K]\:+\:J_x\,\phi_x\:+\:\frac{1}{2}\,K_{xy}\,\big(\phi_x\,\phi_y\:+\:\hbar\,\Delta_{xy}\big)\bigg]\,,
\end{equation}
subject to the extremization
\begin{subequations}
\begin{align}
\frac{\delta \Gamma_{J,K}[\phi,\Delta]}{\delta J_x}\bigg|_{J\,=\,\mathcal{J},\:K\,=\,\mathcal{K}}\ &=\ 0\;,\\\qquad \frac{\delta \Gamma_{J,K}[\phi,\Delta]}{\delta K_{xy}}\bigg|_{J\,=\,\mathcal{J},\:K\,=\,\mathcal{K}}\ &=\ 0\;,
\end{align}
\end{subequations}
where $\mathcal{J}_x$ and $\mathcal{K}_{xy}$ are the \emph{extremal sources}, which we will hereafter refer to as the \emph{external sources}. The generating functional $Z[J,K]$ has the form
\begin{equation}
\label{eq:gen}
Z[J,K] = \int\![\D\Phi]\;\exp\bigg[-\frac{1}{\hbar}\,\bigg(S[\Phi]\:-\: J_x\,\Phi_x\:-\:\frac{1}{2}\,K_{xy}\,\Phi_x\,\Phi_y\bigg)\bigg]\;,
\end{equation}
in which $S[\Phi]$ is the classical Euclidean action. Throughout this article, we employ the DeWitt notation, in which continuous indices are integrated over, i.e.
\begin{equation}
J_x\,\phi_x\ \equiv\ \int\!\D^4x\;J(x)\,\phi(x)\;.
\end{equation}

On performing the extremization in eq.~\eqref{eq:Leg}, we obtain the following expression for the effective action:
\begin{equation}
\label{eq:Leg2}
\Gamma[\phi,\Delta] \ =\ -\,\hbar\,\mathrm{ln}\,Z[\mathcal{J},\mathcal{K}]\:+\:\mathcal{J}_x[\phi,\Delta]\,\phi_x\:+\:\frac{1}{2}\,\mathcal{K}_{xy}[\phi,\Delta]\,\big(\phi_x\,\phi_y\:+\:\hbar\,\Delta_{xy}\big)\,.
\end{equation}
The independent one- and two-point functions, $\phi_x$ and $\Delta_{xy}$, are given by
\begin{subequations}
\label{eq:onetwodef}
\begin{align}
\phi_x\ &=\ \hbar\,\frac{\delta \ln\,Z[J,K]}{\delta J_x}\bigg|_{J\,=\,\mathcal{J},\:K\,=\,\mathcal{K}}\;,\\[0.5em]
\Delta_{xy}\ &=\ 2\,\frac{\delta \ln\,Z[J,K]}{\delta K_{xy}}\bigg|_{J\,=\,\mathcal{J},\:K\,=\,\mathcal{K}}\:-\:\frac{1}{\hbar}\,\phi_x\,\phi_y\;.
\end{align}
\end{subequations}
Note that the sources $J_x$ and $K_{xy}$ are both independent of $\phi_x$ and $\Delta_{xy}$, whereas the external sources $\mathcal{J}_x\equiv \mathcal{J}_x[\phi,\Delta]$ and $\mathcal{K}_{xy}\equiv\mathcal{K}_{xy}[\phi,\Delta]$ are both functionals of $\phi_x$ and $\Delta_{xy}$. As a result, $\phi_x$ and $\Delta_{xy}$ are both functionals of $\mathcal{J}_x$ and $\mathcal{K}_{xy}$, but in such a way that they remain independent of one another. Finally, by means of eq.~\eqref{eq:onetwodef}, we may show that
\begin{subequations}
\label{eq:diffs}
\begin{align}
\label{eq:diffs1}
\frac{\delta \Gamma[\phi,\Delta]}{\delta \phi_x}\ &=\ \mathcal{J}_x[\phi,\Delta]\:+\:\mathcal{K}_{xy}[\phi,\Delta]\,\phi_y\;,\\[0.5em]
\label{eq:diffs2}
\frac{\delta \Gamma[\phi,\Delta]}{\delta \Delta_{xy}}\ &=\ \frac{\hbar}{2}\,\mathcal{K}_{xy}[\phi,\Delta]\;.
\end{align}
\end{subequations}
In the standard approach, eqs.~\eqref{eq:diffs1} and~\eqref{eq:diffs2} deliver the quantum equations of motion for the physical one- and two-point functions when the external sources appearing on the right-hand sides are set to zero. It is in this point and in the way that physical limits will be taken that the present approach will differ.

Before continuing, we remark that the local source $\mathcal{J}_x$ is a real-valued functional. On the other hand, the bi-local source $\mathcal{K}_{xy}$ is, in general, a complex-valued functional. However, it must satisfy the Hermiticity condition $\mathcal{K}_{xy}^*=\mathcal{K}_{yx}$, ensuring that the exponent of the generating functional is real. It follows that the effective action in eq.~\eqref{eq:Leg2} is a real-valued function for $Z[J,K]\geq 0$, acquiring a non-vanishing imaginary part iff $Z[J,K]<0$. The latter situation arises in the context of false vacuum decay, as we will discuss in the following section.

In order to recast the effective action [eq.~\eqref{eq:Leg2}] in a more useful form, we wish to eliminate the explicit appearance of the external sources $\mathcal{J}_x$ and $\mathcal{K}_{xy}$. We must therefore first perform the functional integral in the generating functional $Z[\mathcal{J},\mathcal{K}]$. This can be achieved by expanding around the saddle-point $\varphi_x$, which satisfies the source-dependent stationarity condition
\begin{equation}
\label{eq:deltaS}
\frac{\delta S[\Phi]}{\delta \Phi_x}\bigg|_{\Phi\,=\,\varphi}\:-\:\mathcal{J}_x[\phi,\Delta]\:-\:\mathcal{K}_{xy}[\phi,\Delta]\,\varphi_y \ =\ 0\;.
\end{equation}
Note that the external sources $\mathcal{J}_x$ and $\mathcal{K}_{xy}$, appearing in eq.~\eqref{eq:deltaS}, are evaluated at the configurations $\phi_x$ and $\Delta_{xy}$. By writing $\Phi_x=\varphi_x+\hbar^{1/2}\hat{\Phi}_x$, where the factor of $\hbar^{1/2}$ is included explicitly for bookkeeping purposes, the exponent of the generating functional can be expanded as follows:
\begin{align}
\label{eq:Sexp}
&S[\Phi]\:-\:\mathcal{J}_x[\phi,\Delta]\,\Phi_x\:-\:\frac{1}{2}\,\mathcal{K}_{xy}[\phi,\Delta]\,\Phi_x\,\Phi_y\ =\ S[\varphi]\:-\:\mathcal{J}_x[\phi,\Delta]\,\varphi_x\nonumber\\&\qquad-\:\frac{1}{2}\,\mathcal{K}_{xy}[\phi,\Delta]\,\varphi_x\,\varphi_y\: +\:\frac{\hbar}{2!}\,\hat{\Phi}_x\,\mathcal{G}^{-1}_{xy}[\phi,\Delta]\,\hat{\Phi}_y\:+\:\frac{\hbar^{3/2}}{3!}\,\lambda\,\varphi_x\,\hat{\Phi}_x^3\:+\:\frac{\hbar^2}{4!}\,\lambda\,\hat{\Phi}^4_x\;,
\end{align}
in which there are no terms linear in $\hat{\Phi}_x$ thanks to the stationarity condition in eq.~\eqref{eq:deltaS}. In eq.~\eqref{eq:Sexp}, we have defined the inverse two-point function
\begin{equation}
\label{eq:Gdef}
\mathcal{G}^{-1}_{xy}[\phi,\Delta]\ =\ G_{xy}^{-1}(\varphi)\:-\:\mathcal{K}_{xy}[\phi,\Delta]\;,
\end{equation}
where
\begin{equation}
\label{eq:treeG}
G_{xy}^{-1}(\varphi)\  \equiv\ \frac{\delta^2 S[\Phi]}{\delta\Phi_x\,\delta\Phi_y}\bigg|_{\Phi\,=\,\varphi}\ =\ \frac{\delta^2 S[\varphi]}{\delta\varphi_x\,\delta\varphi_y}\ =\ \delta^{(4)}_{xy} \bigg(\!-\,\partial_x^2\:+\:m^2\:+\:\frac{\lambda}{2}\,\varphi_x^2\bigg)\;,
\end{equation}
and $\delta^{(4)}_{xy}\equiv\delta^{(4)}(x-y)$ is the four-dimensional Dirac delta function. For now, we shall assume that the spectrum of the operator $\mathcal{G}^{-1}_{xy}$ is positive-definite, i.e.~$m^2>0$. We will, however, return to this assumption later in the context of false vacuum decay in section~\ref{sec:CJT}, when we will instead take $m^2<0$.

It is important to recognize that, whereas $\phi_x$ and $\Delta_{xy}$ are independent, $\varphi_x$ and $\mathcal{G}_{xy}$ are not independent by virtue of eq.~\eqref{eq:Gdef}. Since the physical one- and two-point functions necessarily depend upon one-another, it is clear that $\phi_x$ and $\Delta_{xy}$ cannot be physical; only $\varphi_x$ and $\mathcal{G}_{xy}$ can have such an interpretation. 

Substituting the expansion from eq.~\eqref{eq:Sexp} into eq.~\eqref{eq:gen} and performing the functional integral over $\hat{\Phi}_x$, we obtain, at two-loop order, the following expression for the effective action:
\begin{align}
\label{eq:Gammafull}
&\Gamma[\phi,\Delta]\ =\ \Gamma_0[\varphi]\:+\:\hbar\,\Gamma_1[\varphi,\mathcal{G}]\:+\:\hbar^2\,\Gamma_2[\varphi,\mathcal{G}]\:+\:\hbar^2\Gamma_{\mathrm{1PR}}[\varphi,\mathcal{G}]\nonumber\\&\qquad +\,\mathcal{J}_x[\phi,\Delta]\big(\phi-\varphi\big)_x\:+\: \frac{1}{2}\,\mathcal{K}_{xy}[\phi,\Delta]\,\Big[\phi_x\,\phi_y\:-\:\varphi_x\,\varphi_y\:+\:\hbar\,\big(\Delta-\mathcal{G}\big)_{xy}\Big]\;,
\end{align}
where
\begin{subequations}
\begin{align}
\Gamma_0[\varphi]\ &=\ S[\varphi]\;,\\[0.5em]
\label{eq:Gamma1}
\Gamma_1[\varphi,\mathcal{G}]\ &=\ \frac{1}{2}\,\mathrm{tr}\,\Big[\mathrm{ln}\,\big(\mathcal{G}^{-1}\ast G_0\big)\:+\:G^{-1}\ast\mathcal{G}\:-\:1\Big]\nonumber\\[0.5em] &=\ \frac{1}{2}\,\mathrm{tr}\,\Big[\mathrm{ln}\,\big(\mathcal{G}^{-1}\ast G_0\big)\:+\:\mathcal{K}\ast\mathcal{G}\Big]\;,\\[0.5em]
\label{eq:Gamma2}
\hbar^2\,\Gamma_2[\varphi,\mathcal{G}]\ & =\ -\hspace{0.5em}\raisebox{-0.5cm}{\includegraphics[scale=0.75]{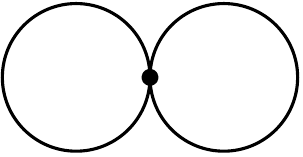}}\hspace{0.5em}-\hspace{0.5em}\raisebox{-0.5cm}{\includegraphics[scale=0.75]{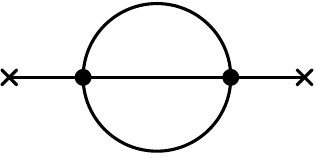}}\;,\\[0.5em]
\label{eq:Gamma1PR}
\hbar^2\,\Gamma_{\mathrm{1PR}}[\varphi,\mathcal{G}]\ &=\ -\hspace{0.5em} \raisebox{-0.5cm}{\includegraphics[scale=0.75]{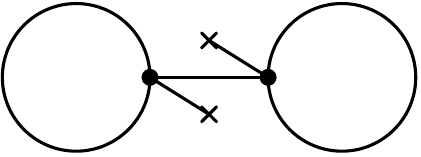}}\;.
\end{align}
\end{subequations}
In eq.~\eqref{eq:Gamma1}, $G_{0,\,xy}\equiv G_{xy}(v)$ is chosen to normalize the fluctuation determinant, where $v$ is some homogeneous vacuum expectation value. In addition, the trace, logarithm and convolution, indicated by an asterisk ($\ast$), are all understood in the functional sense. In eqs.~\eqref{eq:Gamma2} and \eqref{eq:Gamma1PR}, we associate a factor of $\mathcal{G}$ with each line, a factor of $-\,\lambda$ with each vertex and a factor of $\varphi$ with each field insertion. The latter have been represented explicitly for later convenience. In addition, combinatorics have been absorbed; specifically, the double-bubble diagram in eq.~\eqref{eq:Gamma2} and the one-particle-reducible (1PR) dumbbell diagram in eq.~\eqref{eq:Gamma1PR} have combinatorical factors of $1/8$, and the sunset diagram in eq.~\eqref{eq:Gamma2} has a combinatorical factor of $1/12$.  Lastly, our conventions are such that one-loop diagrams are understood to contain an implicit factor of $\hbar$ and two-loop diagrams, an implicit factor of~$\hbar^2$.

In the standard evaluation of the effective action, we would now eliminate the variables $\varphi_x$ and $\mathcal{G}_{xy}$ in favour of $\phi_x$ and $\Delta_{xy}$ (see e.g.~ref.~\cite{Carrington:2004sn}). Instead, we will do the converse, eliminating $\phi_x$ and $\Delta_{xy}$ in favour of $\varphi_x$ and $\mathcal{G}_{xy}$. This alternative expansion was employed in the case of the 1PI effective action in~ref.~\cite{Garbrecht:2015oea} and, in what follows, we will elaborate on the subtleties and merits of this approach within its generalization to include a bi-local external source. We refrain from referring to the present derivation as 2PI, since, as we will see later, this need not be the case in general.

In order to re-express the effective action in terms of $\varphi_x$ and $\mathcal{G}_{xy}$, which we hereafter refer to as the \emph{physical} one- and two-point functions, we proceed in analogy to~ref.~\cite{Carrington:2004sn}. Specifically, we will expand both the left- and right-hand sides of eq.~\eqref{eq:Gammafull} around $\varphi_x=\phi_x-\hbar\,\delta\varphi_x$ and $\mathcal{G}_{xy} = \Delta_{xy}-\hbar\,\delta\mathcal{G}_{xy}$, where
\begin{subequations}
\begin{gather}
\label{eq:deltaphi}
\hbar\,\delta\varphi_x\ =\
\raisebox{-0.45cm}{\includegraphics[scale=0.75]{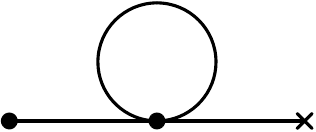}}\;,\\
\label{eq:deltaG}
\hbar\,\delta\mathcal{G}_{xy}\ =\ \raisebox{-0.45cm}{\includegraphics[scale=0.75]{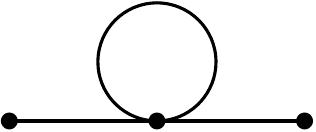}}\hspace{0.5em}+\hspace{0.5em}\raisebox{-0.5cm}{\includegraphics[scale=0.75]{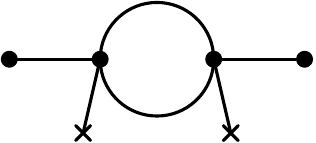}}\hspace{0.5em}+\hspace{0.5em}\raisebox{-0.8cm}{\includegraphics[scale=0.75]{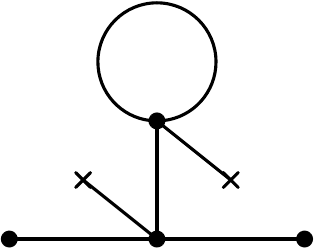}}\;.
\end{gather}
\end{subequations}
The above results can be shown by explicit evaluation of eq.~\eqref{eq:onetwodef} by means of Wick's theorem. In eqs.~\eqref{eq:deltaphi} and~\eqref{eq:deltaG}, lines terminated in a dot indicate untruncated factors of the physical two-point function $\mathcal{G}$.

Expanding the left-hand side of eq.~\eqref{eq:Gammafull} to order $\hbar^2$, we have
\begin{align}
\label{eq:lhsGamma}
\Gamma[\phi,\Delta]\ &=\ \Gamma[\varphi,\mathcal{G}]\:+\:\frac{\delta \Gamma[\phi,\Delta]}{\delta \phi_x}\bigg|_{\varphi,\:\mathcal{G}}\big(\phi-\varphi\big)_x\:+\:\frac{1}{2}\,\frac{\delta^2 \Gamma[\phi,\Delta]}{\delta \phi_x\,\delta\phi_y}\bigg|_{\varphi,\:\mathcal{G}}\big(\phi-\varphi\big)_x\big(\phi-\varphi)_y\nonumber\\&\qquad +\:\frac{\delta \Gamma[\phi,\Delta]}{\delta \Delta_{xy}}\bigg|_{\varphi,\:\mathcal{G}}\big(\Delta-\mathcal{G}\big)_{xy}\:+\:\mathcal{O}(\hbar^3)\;.
\end{align}
Herein, we have used the shorthand subscript ``$\varphi,\:\mathcal{G}$'' to indicate that a term is evaluated at the physical configurations $\phi=\varphi$ and $\Delta=\mathcal{G}$.
Using eq.~\eqref{eq:diffs}, the right-hand side of eq.~\eqref{eq:Gammafull} can be written as
\begin{align}
\label{eq:rhsGamma}
&\Gamma[\phi,\Delta]\ =\ \Gamma_0[\varphi]\:+\:\hbar\,\Gamma_1[\varphi,\mathcal{G}]\:+\:\hbar^2\,\Gamma_2[\varphi,\mathcal{G}]\:+\:\hbar^2\Gamma_{\mathrm{1PR}}[\varphi,\mathcal{G}]\nonumber\\&\qquad +\:\frac{\delta \Gamma[\phi,\Delta]}{\delta \phi_x}\,\big(\phi-\varphi\big)_x\:-\: \frac{1}{\hbar}\,\frac{\delta \Gamma[\phi,\Delta]}{\delta \Delta_{xy}}\,\Big[\big(\phi-\varphi\big)_x\big(\phi-\varphi\big)_y\:-\:\hbar\,\big(\Delta-\mathcal{G}\big)_{xy}\Big]\;.
\end{align}
The functional derivatives appearing on the right-hand side of eq.~\eqref{eq:rhsGamma} are still evaluated at the configurations $\phi$ and $\Delta$, and so we must proceed to expand these around $\varphi$ and $\mathcal{G}$ also. Specifically, we have
\begin{subequations}
\label{eq:diffexps}
\begin{gather}
\frac{\delta \Gamma[\phi,\Delta]}{\delta \phi_x}\ =\ \frac{\delta \Gamma[\phi,\Delta]}{\delta \phi_x}\bigg|_{\varphi,\:\mathcal{G}}\:+\:\frac{\delta^2 \Gamma[\phi,\Delta]}{\delta \phi_x\,\delta\phi_y}\bigg|_{\varphi,\:\mathcal{G}}\big(\phi-\varphi)_y\:+\:\mathcal{O}(\hbar^2)\;,\\[0.5em]
\label{eq:DeltaGammaDelta}
\frac{\delta \Gamma[\phi,\Delta]}{\delta \Delta_{xy}}\ =\ \frac{\delta \Gamma[\phi,\Delta]}{\delta \Delta_{xy}}\bigg|_{\varphi,\:\mathcal{G}}\:+\:\mathcal{O}(\hbar^2)\;.
\end{gather}
\end{subequations}
The first term on the right-hand side of eq.~\eqref{eq:DeltaGammaDelta} is proportional to $\hbar\,\mathcal{K}_{xy}[\varphi,\mathcal{G}]$. As we will confirm in the forthcoming concrete examples, the leading contribution to $\mathcal{K}_{xy}[\varphi,\mathcal{G}]$ is of order $\hbar$. Therefore, we may safely neglect the contribution from eq.~\eqref{eq:DeltaGammaDelta} in eq.~\eqref{eq:rhsGamma} at order $\hbar^2$.  In addition, it can be shown that
\begin{equation}
\label{eq:Gamma_2}
\frac{\delta^2 \Gamma[\phi,\Delta]}{\delta \phi_x\,\delta\phi_y}\bigg|_{\varphi,\:\mathcal{G}}\ =\ \mathcal{G}^{-1}_{xy}[\phi,\Delta]\:+\:\mathcal{O}(\hbar)\;,
\end{equation}
where only the first term on the right-hand side is relevant at the order required here.

Hence, after equating the right-hand sides of eqs.~\eqref{eq:lhsGamma} and~\eqref{eq:rhsGamma}, we substitute eqs.~\eqref{eq:diffexps} and~\eqref{eq:Gamma_2} and cancel like terms, yielding
\begin{align}
\label{eq:Gammamid}
\Gamma[\varphi,\mathcal{G}]\ &=\ \Gamma_0[\varphi]\:+\:\hbar\,\Gamma_1[\varphi,\mathcal{G}]\:+\:\hbar^2\,\Gamma_2[\varphi,\mathcal{G}]\nonumber\\&\qquad +\:\hbar^2\Gamma_{\mathrm{1PR}}[\varphi,\mathcal{G}]\:+\:\frac{1}{2}\,\big(\phi-\varphi)_x\,\mathcal{G}^{-1}_{xy}[\phi,\Delta]\,\big(\phi-\varphi)_y\;.
\end{align}
Finally, using eqs.~\eqref{eq:Gamma1PR} and~\eqref{eq:deltaphi}, we may show that the two terms in the second line of eq.~\eqref{eq:Gammamid} cancel, eliminating the 1PR diagram, as we would anticipate. Thus, we arrive at the final result
\begin{align}
\label{eq:Gammaexp}
\Gamma[\varphi,\mathcal{G}]\ &=\ \Gamma_0[\varphi]\:+\:\hbar\,\Gamma_1[\varphi,\mathcal{G}]\:+\:\hbar^2\,\Gamma_2[\varphi,\mathcal{G}]\;.
\end{align}
Note that the form of eq.~\eqref{eq:Gammaexp} is \emph{identical} to the standard expression for the 2PI effective action with $\phi$ and $\Delta$ replaced by $\varphi$ and $\mathcal{G}$, respectively. The only distinction is that the one- and two-point functions $\varphi$ and $\mathcal{G}$, appearing explicitly in eq.~\eqref{eq:Gammaexp}, \emph{also} appear explicitly in the saddle-point evaluation of the functional integral in the generating functional $Z[\mathcal{J},\mathcal{K}]$. We will return to this point again later in the context of false vacuum decay. What remains to be done is to choose the external sources $\mathcal{J}_x[\phi,\Delta]$ and $\mathcal{K}_{xy}[\phi,\Delta]$ consistently, and we will now proceed to consider three examples that demonstrate the utility of this approach.

\section{CJT 2PI effective action}
\label{sec:CJT}

It is illustrative to recover first the equations of motion corresponding to the usual 2PI effective action due to Cornwall, Jackiw and Tomboulis~\cite{Cornwall:1974vz}. In order to do so, we constrain the form of the external sources $\mathcal{J}_x[\phi,\Delta]$ and $\mathcal{K}_{xy}[\phi,\Delta]$ such that $\varphi_{x}$ and $\mathcal{G}_{xy}$ are the extrema of the effective action, i.e.~we require that
\begin{subequations}
\label{eq:CJTeoms}
\begin{align}
\label{eq:eom1}
\frac{\delta \Gamma[\phi,\Delta]}{\delta \phi_x}\bigg|_{\varphi,\:\mathcal{G}}\ =\ 0\;,\\[0.5em]
\label{eq:eom2}
\frac{\delta \Gamma[\phi,\Delta]}{\delta \Delta_{xy}}\bigg|_{\varphi,\:\mathcal{G}}\ =\ 0\;.
\end{align}
\end{subequations}
We reiterate that $\phi$ and $\Delta$ are independent. It is for this reason that we first take functional derivatives of $\Gamma[\phi,\Delta]$ with respect to $\phi$ and $\Delta$ before taking the limit $\phi\to \varphi$ and $\Delta \to \mathcal{G}$. If, instead, we wished to obtain the correct equations of motion by functionally differentiating $\Gamma[\varphi,\mathcal{G}]$, as it appears in eq.~\eqref{eq:Gammaexp}, with respect to $\varphi$ and $\mathcal{G}$ directly, we would need to define a \emph{partial functional derivative} in order to avoid spurious terms resulting from the mutual dependence of $\varphi$ and $\mathcal{G}$. We do not follow such an approach explicitly in the subsequent analysis.

\paragraph{\bfseries Consistency relation} Equation~\eqref{eq:eom1} gives the quantum equation of motion for the physical one-point function $\varphi_x$, which, at order $\hbar$, takes the form
\begin{equation}
\label{eq:eom1expl}
\frac{\delta \Gamma[\phi,\Delta]}{\delta \phi_x}\bigg|_{\varphi,\mathcal{G}}\ =\ \frac{\delta S[\varphi]}{\delta \varphi_x}\:+\:\hbar\,\frac{\delta \Gamma_1[\phi,\Delta]}{\delta \phi_x}\bigg|_{\varphi,\mathcal{G}}\:+\:\mathcal{O}(\hbar^2)\ =\ 0\;.
\end{equation}
Comparing the right-hand side of eq.~\eqref{eq:eom1expl} with the stationarity condition in eq.~\eqref{eq:deltaS}, we see that (to order $\hbar$)
\begin{equation}
\label{eq:Jconsist}
\frac{\delta S[\varphi]}{\delta \varphi_x}\ =\ \mathcal{J}_{x}[\phi,\Delta]\:+\:\mathcal{K}_{xy}[\phi,\Delta]\,\varphi_y\ 
=\ -\:\hbar\,\frac{\delta \Gamma_1[\phi,\Delta]}{\delta \phi_x}\bigg|_{\varphi,\mathcal{G}}\;.
\end{equation}
We will refer to eq.~\eqref{eq:Jconsist} as the \emph{consistency relation}, which provides one of two constraints on the external sources. As we will see in the remainder of this article, it is through the freedom to choose the other constraint that this method of external sources will acquire its utility. We note also that, were we to go beyond 2PI, we would require additional constraints in order to fix the tri-local and higher sources.

\paragraph{\bfseries 2PI Schwinger-Dyson equation} Equation~\eqref{eq:eom2} gives the Schwinger-Dyson equation for the two-point function $\mathcal{G}_{xy}$, which takes the form
\begin{equation}
\label{eq:SDG}
\mathcal{G}^{-1}_{xy}\ = \ G^{-1}_{xy}\:+\:2\,\hbar\,\frac{\delta \Gamma_2[\phi,\Delta]}{\delta \Delta_{xy}}\bigg|_{\varphi,\mathcal{G}}\;.
\end{equation}
Comparing the right-hand side of eq.~\eqref{eq:SDG} with the definition of the inverse two-point function $\mathcal{G}^{-1}_{xy}$ in eq.~\eqref{eq:Gdef}, it follows that the bi-local external source $\mathcal{K}_{xy}[\phi,\Delta]$ must have the form
\begin{equation}
\label{eq:KCJT}
\mathcal{K}_{xy}[\phi,\Delta]\ =\ -\,2\,\hbar\,\frac{\delta \Gamma_2[\phi,\Delta]}{\delta \Delta_{xy}}\bigg|_{\varphi,\mathcal{G}}\;.
\end{equation}
Substituting eq.~\eqref{eq:KCJT} into the consistency relation in eq.~\eqref{eq:Jconsist}, we find that the local external source $\mathcal{J}_x[\phi,\Delta]$ must have the form
\begin{equation}
\label{eq:JCJT}
\mathcal{J}_x[\phi,\Delta]\ =\ -\:\hbar\,\frac{\delta \Gamma_1[\phi,\Delta]}{\delta \phi_x}\bigg|_{\varphi,\mathcal{G}}\:+\:2\,\hbar\,\frac{\delta \Gamma_2[\phi,\Delta]}{\delta \Delta_{xy}}\bigg|_{\varphi,\mathcal{G}}\,\varphi_y\;.
\end{equation}
We see from eqs.~\eqref{eq:KCJT} and~\eqref{eq:JCJT} that the external sources are both formally order $\hbar$, and it was indeed correct to neglect the contribution to the effective action in eq.~\eqref{eq:rhsGamma} from eq.~\eqref{eq:DeltaGammaDelta} at orders up to and including $\hbar^2$.

We remark that the one-particle-irreducible (1PI) effective action~\cite{Jackiw:1974cv} may be recovered straightforwardly in this approach. Specifically, in the limit that $\mathcal{K}_{xy}[\phi,\Delta]$ is taken to zero, it is trivially the case from its definition in eq.~\eqref{eq:Gdef} that $\mathcal{G}_{xy}^{-1} = G_{xy}^{-1}$. The consistency relation in eq.~\eqref{eq:Jconsist} then reduces to that imposed upon the local source $\mathcal{J}_x[\phi,\Delta]$ in ref.~\cite{Garbrecht:2015oea}, where the alternative evaluation elaborated upon here was applied to the 1PI effective action in the context of false vacuum decay.

The self-consistency of this method of evaluating the effective action requires the identities quoted in eq.~\eqref{eq:CJTeoms} to hold. This can be the case only if the external sources $\mathcal{J}_x[\phi,\Delta]$ and $\mathcal{K}_{xy}[\phi,\Delta]$ in eqs.~\eqref{eq:KCJT} and~\eqref{eq:JCJT} vanish when evaluated at the physical one- and two-point configurations $\varphi_x$ and $\mathcal{G}_{xy}$, i.e.
\begin{subequations}
\label{eq:source0}
\begin{align}
\label{eq:J0}
\mathcal{J}_x[\varphi,\mathcal{G}]\ &=\ 0\;,\\[0.5em]
\label{eq:K0}
\mathcal{K}_{xy}[\varphi,\mathcal{G}]\ &=\ 0\;.
\end{align}
\end{subequations}
In order to show that this is indeed the case, we must expand the functional arguments of the external sources $\mathcal{J}_x[\phi,\Delta]$ and $\mathcal{K}_{xy}[\phi,\Delta]$ around $\varphi_x$ and $\mathcal{G}_{xy}$ to order $\hbar$.

To this end, we first consider the combination of external sources appearing in eq.~\eqref{eq:diffs1}, which may be expanded as follows:
\begin{align}
\label{eq:Jexp}
&\mathcal{J}_x[\phi,\Delta]\:+\:\mathcal{K}_{xy}[\phi,\Delta]\phi_y\ =\ \mathcal{J}_x[\varphi,\mathcal{G}]\:+\:\mathcal{K}_{xy}[\varphi,\mathcal{G}]\varphi_y\nonumber\\&\qquad +\;\frac{\delta^2\Gamma[\phi,\Delta]}{\delta \phi_x\,\delta\phi_y}\bigg|_{\varphi,\:\mathcal{G}}\big(\phi-\varphi)_y\:+\:\frac{\delta^2\Gamma[\phi,\Delta]}{\delta\phi_x\,\delta \Delta_{yz}}\bigg|_{\varphi,\:\mathcal{G}}\big(\Delta-\mathcal{G}\big)_{yz}\:+\:\cdots
\end{align}
By virtue of eq.~\eqref{eq:Gamma_2}, the first term in the second line of eq.~\eqref{eq:Jexp} gives
\begin{equation}
\hbar\,\mathcal{G}^{-1}_{xy}[\phi,\Delta]\delta\varphi_y\ =\ -\:\hbar\,\frac{\delta \Gamma_1[\phi,\Delta]}{\delta \phi_x}\bigg|_{\varphi,\mathcal{G}}\ = \hspace{0.5em}\raisebox{-0.4cm}{\includegraphics[scale=0.75]{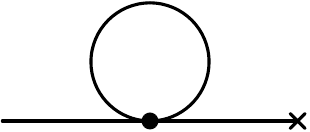}}\hspace{0.5em}\;,
\end{equation}
where $\delta\varphi_x$ is defined in eq.~\eqref{eq:deltaphi}. Since $\phi$ and $\Delta$ are independent, the second term in the second line of eq.~\eqref{eq:Jexp} is zero at order $\hbar$; specifically,
\begin{equation}
\frac{\delta^2\Gamma[\phi,\Delta]}{\delta\phi_x\,\delta \Delta_{yz}}\bigg|_{\varphi,\:\mathcal{G}}\big(\Delta-\mathcal{G}\big)_{yz}\ =\ \frac{\delta^2\Gamma_0[\phi]}{\delta\phi_x\,\delta \Delta_{yz}}\bigg|_{\varphi,\:\mathcal{G}}\big(\Delta-\mathcal{G}\big)_{yz}\:+\:\mathcal{O}(\hbar^2)\ =\ \mathcal{O}(\hbar^2)\;.
\end{equation}
Thus, we have
\begin{align}
\label{eq:Jexp2}
&\mathcal{J}_x[\phi,\Delta]\:+\:\mathcal{K}_{xy}[\phi,\Delta]\phi_y\ =\ \mathcal{J}_x[\varphi,\mathcal{G}]\:+\:\mathcal{K}_{xy}[\varphi,\mathcal{G}]\varphi_y\:-\:\hbar\,\frac{\delta \Gamma_1[\phi,\Delta]}{\delta \phi_x}\bigg|_{\varphi,\mathcal{G}}\;.
\end{align}
Since we may write
\begin{align}
\mathcal{J}_x[\phi,\Delta]\:+\:\mathcal{K}_{xy}[\phi,\Delta]\phi_y\ &= \ \mathcal{J}_x[\phi,\Delta]\:+\:\mathcal{K}_{xy}[\phi,\Delta](\varphi_y\:+\:\hbar\,\delta\varphi_y)\nonumber\\ &=\ \mathcal{J}_x[\phi,\Delta]\:+\:\mathcal{K}_{xy}[\phi,\Delta]\varphi_y\:+\:\mathcal{O}(\hbar^2)\;,
\end{align}
it follows, by comparing eq.~\eqref{eq:Jexp2} with eq.~\eqref{eq:Jconsist}, that
\begin{equation}
\label{eq:JK0}
\mathcal{J}_x[\varphi,\mathcal{G}]\:+\:\mathcal{K}_{xy}[\varphi,\mathcal{G}]\,\varphi_y\ =\ 0
\end{equation}
at order $\hbar$.

Next, we  expand the bi-local source $\mathcal{K}_{xy}[\phi,\Delta]$ in isolation, giving
\begin{equation}
\mathcal{K}_{xy}[\phi,\Delta]\ =\ \mathcal{K}_{xy}[\varphi,\mathcal{G}]\:+\:\frac{\delta\mathcal{K}_{xy}[\phi,\Delta]}{\delta \phi_z}\,\bigg|_{\varphi,\mathcal{G}}\big(\phi-\varphi\big)_z\:+\:\frac{\delta\mathcal{K}_{xy}[\phi,\Delta]}{\delta \Delta_{zw}}\,\bigg|_{\varphi,\mathcal{G}}\big(\Delta-\mathcal{G}\big)_{zw}\:+\:\cdots\;.
\end{equation}
Using eq.~\eqref{eq:diffs2}, this expansion may be re-expressed in terms of functional derivatives of the effective action as
\begin{equation}
\label{eq:Kexp}
\mathcal{K}_{xy}[\phi,\Delta]\ =\ \mathcal{K}_{xy}[\varphi,\mathcal{G}]\:+\:\frac{2}{\hbar}\,\frac{\delta^2\Gamma[\phi,\Delta]}{\delta \Delta_{xy}\, \delta \phi_z}\bigg|_{\varphi,\:\mathcal{G}}\big(\phi-\varphi\big)_z\:+\:\frac{2}{\hbar}\,\frac{\delta^2\Gamma[\phi,\Delta]}{\delta \Delta_{xy}\, \delta \Delta_{zw}}\bigg|_{\varphi,\:\mathcal{G}}\big(\Delta-\mathcal{G}\big)_{zw}\:+\:\cdots\;.
\end{equation}
The terms of zeroth order in $\hbar$ are vanishing by virtue of the independence of $\phi$ and $\Delta$, and of interest to us are the terms of order $\hbar$, namely
\begin{equation}
\label{eq:Kdiagexp}
\mathcal{K}_{xy}[\phi,\Delta]\ \supset\ 2\,\hbar\,\frac{\delta^2\Gamma_1[\phi,\Delta]}{\delta \Delta_{xy}\, \delta \phi_z}\bigg|_{\varphi,\:\mathcal{G}}\delta\varphi_z\:+\:2\,\hbar\,\frac{\delta^2\Gamma_1[\phi,\Delta]}{\delta \Delta_{xy}\, \delta \Delta_{zw}}\bigg|_{\varphi,\:\mathcal{G}}\delta\mathcal{G}_{zw}\;.
\end{equation}
Proceeding diagrammatically, we can show that
\begin{subequations}
\begin{gather}
\label{eq:lolli1}
2\,\hbar\,\frac{\delta^2\Gamma_1[\phi,\Delta]}{\delta \Delta_{xy}\, \delta \phi_z}\bigg|_{\varphi,\:\mathcal{G}}\delta\varphi_z\ =\ -\hspace{0.5em}\raisebox{-2em}{\includegraphics[scale=0.75]{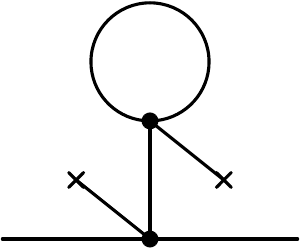}}\hspace{0.5em},\\
\label{eq:lolli2}
2\,\hbar\,\frac{\delta^2\Gamma_1[\phi,\Delta]}{\delta \Delta_{xy}\, \delta \Delta_{zw}}\bigg|_{\varphi,\:\mathcal{G}}\delta\mathcal{G}_{zw}\ =\  \raisebox{-0.45cm}{\includegraphics[scale=0.75]{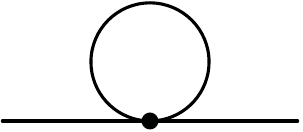}}\hspace{0.5em}+\hspace{0.5em}\raisebox{-0.52cm}{\includegraphics[scale=0.75]{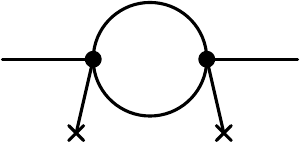}}\hspace{0.5em}+\hspace{0.5em}\raisebox{-2em}{\includegraphics[scale=0.75]{lollipop.pdf}}\hspace{0.5em}.
\end{gather}
\end{subequations}
Hence, in eq.~\eqref{eq:Kdiagexp}, the lollipop diagrams in eqs.~\eqref{eq:lolli1} and~\eqref{eq:lolli2} cancel, and we are left with
\begin{equation}
\label{eq:lastKrel}
2\,\hbar\,\frac{\delta^2\Gamma_1[\phi,\Delta]}{\delta \Delta_{xy}\, \delta \phi_z}\bigg|_{\varphi,\:\mathcal{G}}\delta\varphi_z\:+\:2\,\hbar\,\frac{\delta^2\Gamma_1[\phi,\Delta]}{\delta \Delta_{xy}\, \delta \Delta_{zw}}\bigg|_{\varphi,\:\mathcal{G}}\delta\mathcal{G}_{zw}\ =\ -\:2\,\hbar\,\frac{\delta\Gamma_2[\phi,\Delta]}{\delta \Delta_{xy}}\bigg|_{\varphi,\mathcal{G}}\;.
\end{equation}
Comparing this result with eqs.~\eqref{eq:KCJT} and~\eqref{eq:JK0}, both eqs.~\eqref{eq:J0} and~\eqref{eq:K0} follow directly. Therefore, we have shown that
\begin{subequations}
\begin{gather}
\frac{\delta \Gamma[\phi,\Delta]}{\delta \phi_x}\bigg|_{\varphi,\:\mathcal{G}}\ =\ \bigg[\mathcal{J}_x[\phi,\Delta]\:+\:\mathcal{K}_{xy}[\phi,\Delta]\phi_y\bigg]_{\varphi,\:\mathcal{G}}\ =\ 0\;,\\[0.5em]
\frac{\delta \Gamma[\phi,\Delta]}{\delta \Delta_{xy}}\bigg|_{\varphi,\:\mathcal{G}}\ = \ \bigg[\frac{\hbar}{2}\,\mathcal{K}_{xy}[\phi,\Delta]\bigg]_{\varphi,\:\mathcal{G}}\ =\ 0\;,
\end{gather}
\end{subequations}
and established the identities in eq.~\eqref{eq:CJTeoms} at order $\hbar$, as required.

In summary, we have seen that by constraining the external sources such that the physical one- and two-point functions $\varphi_x$ and $\mathcal{G}_{xy}$ are the extrema of the effective action, we are able to recover the standard 2PI CJT effective action. There is, however, one significant difference from the standard result: the stationarity condition in eq.~\eqref{eq:deltaS} means that the functional integral in the generating functional $Z[\mathcal{J},\mathcal{K}]$ is evaluated about the extremal \emph{quantum} path, not the extremal \emph{classical} path, as would be the case in the standard approach. This latter fact is of particular relevance to situations in which the extremal quantum path is non-perturbatively far away from the extremal classical path. An example of this occurs in the case of vacuum decay when the global minimum of the potential is generated only by radiative effects, as we will discuss below.

\paragraph{\bfseries Relevance to vacuum decay} For systems exhibiting vacuum instability, the tunneling rate for transitions between false and true vacua may be calculated from the Euclidean path integral by expanding around the so-called ``bounce'' configuration (see e.g.~the seminal works by Coleman and Callan~\cite{Coleman:1977py,Callan:1977pt}). When the instability arises at tree-level, the bounce is the solution to the classical equation of motion that interpolates between false and true vacua. In other words, the bounce looks like a four-dimensional hyperspherical bubble, which separates true vacuum on the inside from false vacuum on the outside.

For definiteness, let us consider the archetypal case of the $\lambda\Phi^4$ theory with tachyonic mass $m^2<0$ [see eq.~\eqref{eq:Lag}]. In this case, working in hyperspherical coordinates, the bounce is given by the well-known kink solution
\begin{equation}
\varphi(r)\ = \ v\,\tanh[\gamma(r-R)]\;,
\end{equation}
where $v=\sqrt{12\gamma^2/\lambda}$ is the vacuum expectation value at the global minimum of the potential, $\gamma=m/\sqrt{2}$ and $R$ is the radius of the critical bubble.\footnote{In order to obtain a bubble with finite radius $R$, we must provide some small breaking of the $\mathbb{Z}_2$ symmetry of the Lagrangian in eq.~\eqref{eq:Lag} and thereby of the degeneracy of the two minima in the potential. This can be achieved, for instance, by adding a cubic term of the form $g\Phi^3/3!$, in which case $R\sim 1/g$.} The tunneling rate per unit volume is given by
\begin{equation}
\label{eq:tunnrate}
\varGamma/V\ =\ 2\,\big|\mathrm{Im}\,Z[0]\big|/(VT)\;,
\end{equation}
where $V$ is the volume and $T$ is the Euclidean time of the bounce, and
\begin{equation}
\label{eq:Z0}
Z[0]\ =\ \int\![\D\Phi]\;e^{-S[\Phi]/\hbar}
\end{equation}
is the Euclidean path integral. This path integral, although seemingly real-valued, acquires a non-zero imaginary part as a result of the instability. In order to see this, we perform a saddle-point evaluation of the functional integral in eq.~\eqref{eq:Z0} by expanding around the \emph{classical} bounce, writing $\varphi=\Phi-\hbar^{1/2}\hat{\Phi}$. In this way, we obtain
\begin{equation}
\label{eq:Z0saddle}
Z[0]\ =\ e^{-S[\varphi]/\hbar}\int[\mathrm{d}\hat{\Phi}]\;e^{-\,\hat{\Phi}_xG_{xy}^{-1}(\varphi)\hat{\Phi}_y}\;,
\end{equation}
where the Klein-Gordon operator $G_{xy}^{-1}(\varphi)$, as given by eq.~\eqref{eq:treeG}, is proportional to the tree-level fluctuation operator
\begin{equation}
\mathcal{M}_x(\varphi)\ \equiv\ -\:\partial^2_x\:+\:m^2\:+\:\frac{\lambda}{2}\,\varphi_x^2\;.
\end{equation}
For the case $m^2<0$, this operator acquires a non-positive-definite spectrum. Specifically, the fluctuation operator has four zero eigenvalues, corresponding to the translational invariance of the bounce action $S[\varphi]$, and one negative eigenvalue, corresponding to dilatations of the bounce. As a result, the Gaussian functional integral in eq.~\eqref{eq:Z0saddle} is ill defined. Even so, the functional integral over the zero modes can be traded for an integral over the collective coordinates of the bounce, giving rise to a spacetime volume prefactor $VT$ in the tunneling rate. The presence of the negative eigenvalue requires us to deform the contour of integration into the complex plane by means of the method of steepest descent, giving rise to the non-zero imaginary part featuring in eq.~\eqref{eq:tunnrate}.

The issue of false vacuum decay becomes less straightforward when the instability arises purely from radiative effects. In this case, the tree-level fluctuation operator will have a positive-definite spectrum, whilst the perturbatively-calculated effective potential becomes non-convex, indicating the presence of the instability~\cite{Weinberg:1987vp}. However, since, in the absence of external sources, it is the tree-level fluctuation operator that arises in the saddle-point evaluation of eq.~\eqref{eq:Z0}, the origin of the non-zero imaginary part and the correct approach for determining the tunneling rate is less clear.

The situation is somewhat different when we consider the alternative method of evaluating the effective action presented above. The relevant path integral has the form
\begin{equation}
Z[\mathcal{J},\mathcal{K}]\ =\ \int[\mathrm{d}\Phi]\,e^{-(S[\Phi]-\mathcal{J}_x\Phi_x-\frac{1}{2}\Phi_x\mathcal{K}_{xy}\Phi_y)/\hbar}\;,
\end{equation}
where $\mathcal{K}_{xy}\equiv \mathcal{K}_{xy}[\phi,\Delta]$ and $\mathcal{J}_x\equiv\mathcal{J}_x[\phi,\Delta]$ are the non-vanishing external sources, as given respectively by eqs.~\eqref{eq:KCJT} and \eqref{eq:JCJT}. In order to find the tunneling rate when the instability arises beyond tree-level, say at one-loop level, we wish to expand the functional integral around the \emph{quantum} bounce, which is the solution to eq.~\eqref{eq:eom1}. Proceeding in this manner, writing $\varphi=\Phi-\hbar^{1/2}\hat{\Phi}$ as before, we obtain
\begin{equation}
\label{eq:ZJKexp}
Z[\mathcal{J},\mathcal{K}]\ =\ e^{-S[\varphi]/\hbar} \int[\mathrm{d}\hat{\Phi}]\,e^{-\,\hat{\Phi}_x(G^{-1}_{xy}(\varphi)\:-\:\mathcal{K}_{xy})\hat{\Phi}_y}\;.
\end{equation}
We emphasize that $\mathcal{K}_{xy}$ is still evaluated at the configurations $\phi$ and $\Delta$ in eq.~\eqref{eq:ZJKexp}.
By comparing the exponent in eq.~\eqref{eq:ZJKexp} with eq.~\eqref{eq:Gdef}, we see that the kernel of the Gaussian integral is now the \emph{dressed} inverse two-point function $\mathcal{G}_{xy}^{-1}[\phi,\Delta]$. By virtue of eq.~\eqref{eq:KCJT}, this operator contains the one-loop corrections. Thus, if the vacuum instability is generated at the one-loop level, $\mathcal{G}_{xy}^{-1}[\phi,\Delta]$ will have a non-positive-definite spectrum, comprising the four zero eigenvalues and one negative eigenvalue in complete analogy to the case in which the instability arises at tree-level. As a result, we may straightforwardly relate the tunneling rate per unit volume to the imaginary part of the effective action via
\begin{equation}
\varGamma/V\ =\ 2|\mathrm{Im}\,e^{-\Gamma[\varphi,\mathcal{G}]/\hbar}|/(VT)\;,
\end{equation}
where this non-zero imaginary part arises again from having necessarily to employ the method of steepest descent in order to deal with the presence of the negative eigenvalue.

We note that, in the generic scenario considered above, the extremal classical and quantum paths are non-perturbatively far away from one another: the first corresponds to a constant \emph{homogeneous} background field configuration with a positive-definite spectrum of quadratic fluctuations, and the second corresponds instead to an \emph{inhomogeneous} background field configuration with a non-positive-definite spectrum of fluctuations. A concrete example, where the one-loop corrections induce a non-convex region in the effective potential that is not present at tree-level, is the Coleman-Weinberg mechanism of SSB, to which the present method was applied in the context of vacuum decay in ref.~\cite{Garbrecht:2015yza}.

\section{CV 2PPI effective action}
\label{sec:VC}

In this section, we will consider an example in which the bi-local source is chosen so as to yield a different variant of the effective action. Specifically, we will recover the CV two-particle-point-irreducible (2PPI) effective action~\cite{Verschelde:1992bs, Verschelde:1992ig, Coppens:1993zc, Coppens:1993ri}. This is defined by the Legendre transform
\begin{equation}
\Gamma^{\rm 2PPI}[\phi,\Delta]\ =\ -\,\hbar\,\ln\,Z[\mathcal{J},\mathcal{K}]\:+\:\mathcal{J}_x[\phi,\Delta]\:+\:\frac{1}{2}\,\mathcal{K}_{x}[\phi,\Delta]\,\big(\phi_x^2\:+\:\hbar\,\Delta_{xx}\big)\;.
\end{equation}
The CV 2PPI effective action differs from the CJT 2PI effective action in that the sources $\mathcal{J}_x[\phi,\Delta]$ and $\mathcal{K}_x[\phi,\Delta]$ are both \emph{local}, respectively constraining the expectation values of $\Phi_x$ and $\Phi_x^2$ rather than $\Phi_x$ and $\Phi_x\Phi_y$. After eliminating $\mathcal{J}_x[\phi,\Delta]$ and $\mathcal{K}_x[\phi,\Delta]$ in favour of $\phi_x$ and $\Delta_{xx}$, the 2PPI effective action takes the form\footnote{Equation~\eqref{eq:G2PPI} is understood in the notation of the standard approach to the effective action, wherein $\phi$ and $\Delta$ are interpreted as the \emph{physical} one- and two-point functions and are no longer independent.}~\cite{Verschelde:1992bs, Verschelde:1992ig, Coppens:1993zc, Coppens:1993ri} 
\begin{equation}
\label{eq:G2PPI}
\Gamma^{\rm 2PPI}[\phi,\Delta]\ =\ S[\phi]\:+\:\hbar\,\Gamma^{\rm 2PPI}_1[\phi,M^2(\phi,\Delta)]\:+\:\hbar^2\,\Gamma_2^{\rm 2PPI}[\phi,M^2(\phi,\Delta)]\:-\:\frac{\lambda}{8}\,\hbar^2\,\Delta_{xx}^2\;,
\end{equation}
where $M(\phi,\Delta)$ is the effective mass, given by
\begin{equation}
M^2(\phi,\Delta)\ =\ m^2\:+\:\frac{\lambda}{2}\,\big(\phi^2\:+\:\hbar\,\Delta_{xx}\big)\;.
\end{equation}
In addition, the one- and two-loop corrections of the 2PPI effective action are
\begin{subequations}
\begin{align}
\label{eq:G12PPI}
\hbar\,\Gamma_1^{\rm 2PPI}[\phi,M^2(\phi,\Delta)]\ &=\ \frac{\hbar}{2}\,\mathrm{tr}\,\mathrm{ln}\,\big(\Delta^{-1}\ast G_0\big)\;,\\
\label{eq:G22PPI}
\hbar^2\,\Gamma_2^{\rm 2PPI}[\phi,M^2(\phi,\Delta)]\ &=\ -\hspace{0.5em}\raisebox{-0.5cm}{\includegraphics[scale=0.75]{2PI2.pdf}}\;,
\end{align}
\end{subequations}
where
\begin{equation}
\label{eq:VCSD}
\Delta^{-1}_{xy}\ =\ \delta^{(4)}_{xy}\big(-\partial_x^2\:+\:M^2(\phi,\Delta)\big)
\end{equation}
and we associate a factor of $\Delta_{xy}$ with each diagrammatic line. The diagram in eq.~\eqref{eq:G22PPI} is the only two-loop 2PPI diagram. The latter is defined to be a 1PI diagram that stays connected when two internal lines meeting at the same vertex are cut open. The double-bubble diagram, present in eq.~\eqref{eq:Gamma2} for the 2PI effective action, is not 2PPI and has been isolated as the last term in eq.~\eqref{eq:G2PPI}. Notice, however, that it has the wrong sign compared to the 2PI case. As we will see in what follows, this is due to the fact that there are terms missing in eq.~\eqref{eq:G12PPI} compared to eq.~\eqref{eq:Gamma1}. These missing terms also provide additional two-loop diagrams.

We can identify the two-point function $\Delta_{xx}$ at coincidence with 
\begin{align}
\label{eq:Dxx}
\Delta_{xx}\ &=\ 2\,\frac{\delta\Gamma_1^{\rm 2PPI}[\phi,M^2(\phi,\Delta)]}{\delta M^2(\phi,\Delta)}\:+\:2\,\hbar\,\frac{\delta\Gamma_2^{\rm 2PPI}[\phi,M^2(\phi,\Delta)]}{\delta M^2(\phi,\Delta)}\nonumber\\[0.5em] &=\ -\:\frac{2}{\hbar\,\lambda}\,\Bigg[\hspace{0.5em}\raisebox{-0.45cm}{\includegraphics[scale=0.75]{self1.pdf}}\hspace{0.5em}+\hspace{0.5em}\raisebox{-0.4cm}{\includegraphics[scale=0.75]{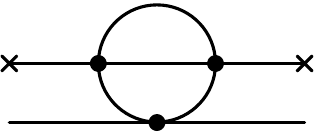}}\hspace{0.5em}\Bigg]\;.
\end{align}
The functional derivatives have been performed by using the functional chain rule and the fact that
\begin{equation}
\label{eq:diagdif}
\frac{\delta\Delta_{xy}}{\delta M^2(\phi,\Delta)}\ =\ -\:\Delta_{xz}\,\Delta_{zy}\:+\:\mathcal{O}(\hbar)\ =\ \frac{1}{\lambda}\hspace{0.5em}\raisebox{-0.7em}{\includegraphics[scale=0.75]{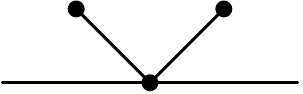}}\hspace{0.5em}+\:\mathcal{O}(\hbar)\;.
\end{equation}
Hence, from eqs.~\eqref{eq:VCSD} and~\eqref{eq:Dxx}, we see that the 2PPI approach resums all point insertions of local self-energies in the propagator $\Delta_{xy}$. Note that, were the double-bubble diagram to have been included in $\Gamma_2^{\rm 2PPI}[\phi,M^2(\phi,\Delta)]$, we would also have generated the diagram
\begin{equation}
-\,\frac{\lambda}{2}\,\hbar^2\,\Delta_{xx}\,\frac{\delta \Delta_{xx}}{\delta M^2(\phi,\Delta)}\ =\ \frac{2}{\lambda}\,\hspace{0.5em}\raisebox{-2em}{\includegraphics[scale=0.75]{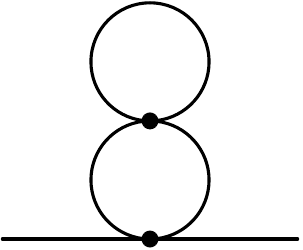}}\hspace{0.5em}\;,
\end{equation}
thereby double-counting the resummation of the one-loop tadpole insertions, already included in the first diagram of eq.~\eqref{eq:Dxx}.

We will now show how the 2PPI effective action may readily be recovered from eq.~\eqref{eq:Gammaexp} in this alternative approach by appropriate choice of the bi-local source $\mathcal{K}_{xy}[\phi,\Delta]$. Moreover, it will not be necessary to isolate the double-bubble diagram in order to avoid double-counting of the diagrammatic series. In the first instance, we will work at the one-loop level, truncating the Schwinger-Dyson equation and thereby the local and bi-local sources at order $\hbar$. The one-loop result will then allow us to obtain the two-loop result iteratively.

For the 2PPI effective action, we must still require that the physical one-point function $\varphi$ is the extremum of the effective action, i.e.
\begin{equation}
\label{eq:2PPIonshell}
\frac{\delta \Gamma[\phi,\Delta]}{\delta \phi_x}\bigg|_{\varphi,\mathcal{G}}\ =\ \mathcal{J}_x[\varphi,\mathcal{G}]\:+\:\mathcal{K}_{xy}[\varphi,\mathcal{G}]\varphi_y\ =\ 0\;.
\end{equation}
As such, the external sources are still subject to the consistency relation in eq.~\eqref{eq:Jconsist}. However, it will not be the case that $\mathcal{G}_{xy}$ is the extremal two-point function, i.e.~we will find that
\begin{equation}
\label{eq:G2neq}
\frac{\delta \Gamma[\phi,\Delta]}{\delta \Delta_{xy}}\bigg|_{\varphi,\mathcal{G}}\ \neq \ 0
\end{equation}
and therefore that $\mathcal{G}_{xy}$ is not the solution to the 2PI Schwinger-Dyson equation in eq.~\eqref{eq:SDG}. As we will show explicitly, the appropriate choice for the bi-local source is
\begin{equation}
\label{eq:VCK}
\mathcal{K}_{xy}[\phi,\Delta]\ =\ -\:2\hbar\,\frac{\delta \Gamma_{1}[\varphi,\mathcal{G}]}{\delta \varphi_x^2}\,\delta^{(4)}_{xy}\:+\:\mathcal{O}(\hbar^2)\;.
\end{equation}
Here, it is important to emphasize the following two points: (i) we take the functional derivative of $\Gamma_1[\varphi,\mathcal{G}]$ and not $\Gamma_1[\phi,\Delta]$, such that we must account for the fact that $\varphi$ and $\mathcal{G}$ are not independent; and (ii) the functional derivative in eq.~\eqref{eq:VCK} is taken with respect to the \emph{square} of the physical one-point function. By substituting this form for the bi-local source into the equation for the inverse two-point function $\mathcal{G}^{-1}_{xy}$ in eq.~\eqref{eq:Gdef}, it is clear that we resum only the local tapdole insertions. However, in order to verify that we do indeed recover the 2PPI effective action, it is helpful to go beyond the one-loop level. In particular, given that the one-loop diagram in eq.~\eqref{eq:VCK} is itself a functional of the source $\mathcal{K}_{xy}[\phi,\Delta]$, one might worry about the consistency of this naive $\hbar$ truncation.

\paragraph{\bfseries 2PPI Schwinger-Dyson equation} Before proceeding to the two-loop calculation, it is illustrative to consider the consistency relation in eq.~\eqref{eq:Jconsist} and the form of the local source $\mathcal{J}_{x}[\phi,\Delta]$. From the former, we have the constraint
\begin{equation}
\mathcal{J}_x[\phi,\Delta]\:+\:\mathcal{K}_{xy}[\phi,\Delta]\varphi_y\ =\ \hspace{0.5em}\raisebox{-0.45cm}{\includegraphics[scale=0.75]{tad.pdf}}\hspace{0.5em}\;.
\end{equation}
Substituting for $\mathcal{K}_{xy}[\phi,\Delta]$ from eq.~\eqref{eq:VCK}, it immediately follows that
\begin{equation}
\mathcal{J}_x[\phi,\Delta]\ =\ 0
\end{equation}
at order $\hbar$. In order to find the explicit form of eq.~\eqref{eq:G2neq}, we proceed to expand $\mathcal{K}_{xy}[\phi,\Delta]$ around the physical one- and two-point functions $\varphi$ and $\mathcal{G}$, following the same arguments as in section~\ref{sec:CJT} [cf.~eqs.~\eqref{eq:Kexp}--\eqref{eq:lastKrel}]. In this way, we obtain
\begin{equation}
\mathcal{K}_{xy}[\varphi,\mathcal{G}]\ =\ -\hspace{0.5em}\raisebox{-0.5cm}{\includegraphics[scale=0.75]{self2.pdf}}\hspace{0.5em}\;.
\end{equation}
By virtue of eq.~\eqref{eq:2PPIonshell}, it then follows that
\begin{equation}
\mathcal{J}_{x}[\varphi,\mathcal{G}]\ = \hspace{0.5em}\raisebox{-0.52cm}{\includegraphics[scale=0.75]{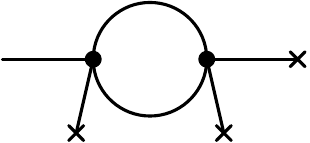}}\hspace{0.5em}\;.
\end{equation}
Finally, the Schwinger-Dyson equation reads [cf.~eqs.~\eqref{eq:diffs2} and~\eqref{eq:G2neq}]
\begin{equation}
\frac{\delta \Gamma[\phi,\Delta]}{\delta \Delta_{xy}}\bigg|_{\varphi,\mathcal{G}}\ =\ -\:\frac{\hbar}{2}\,\mathcal{G}^{-1}_{xy}\:+\:\frac{\hbar}{2}\,G^{-1}_{xy}\:+\:\hbar^2\,\frac{\delta \Gamma_2[\phi,\Delta]}{\delta \Delta_{xy}}\bigg|_{\varphi,\mathcal{G}}\ =\ -\:\frac{\hbar}{2}\,\hspace{0.25em}\raisebox{-0.5cm}{\includegraphics[scale=0.75]{self2.pdf}}\hspace{0.5em}\;.
\end{equation}
This result is as we would expect: the non-vanishing right-hand side is such that it cancels the non-local bubble diagram in the 2PI self-energy, which arises from
\begin{equation}
\hbar^2\frac{\delta \Gamma_2[\phi,\Delta]}{\delta \Delta_{xy}}\bigg|_{\varphi,\mathcal{G}}\ =\ -\,\frac{\hbar}{2}\bigg[\hspace{0.5em}\raisebox{-0.45cm}{\includegraphics[scale=0.75]{self1.pdf}}\hspace{0.5em}+\hspace{0.25em}\raisebox{-0.5cm}{\includegraphics[scale=0.75]{self2.pdf}}\hspace{0.5em}\bigg]\;,
\end{equation}
leaving behind only the local tadpole self-energy.

In order to check the consistency of the naive truncation of the $\hbar$ expansion employed above, we will now analyze this method of formulating the 2PPI effective action at the two-loop level. To this end, we first look at the ``one-loop'' term $\Gamma_1[\varphi,\mathcal{G}]$ in the light of the results above. From eq.~\eqref{eq:Gamma1}, we have
\begin{equation}
\label{eq:Gamma1K}
\hbar\,\Gamma_1[\varphi,\mathcal{G}]\ =\ \frac{\hbar}{2}\,\mathrm{tr}\,\Big[\mathrm{ln}\,\big(\mathcal{G}^{-1}\ast G_0\big)\:+\:\mathcal{K}\ast\mathcal{G}\Big]\;.
\end{equation}
Given the one-loop result for the bi-local source $\mathcal{K}_{xy}[\phi,\Delta]$ in eq.~\eqref{eq:VCK}, we see that the second term in eq.~\eqref{eq:Gamma1K} has the following diagrammatic form:
\begin{equation}
\label{eq:Gamma1bub}
\hbar\,\Gamma_1[\varphi,\mathcal{G}]\ \supset\ \frac{\hbar}{2}\,\mathrm{tr}\,\mathcal{K}\ast\mathcal{G}\ =\ 2 \hspace{0.5em}\raisebox{-0.45cm}{\includegraphics[scale=0.75]{2PI1.pdf}}\hspace{0.5em}\;.
\end{equation}
It is the appearance of this additional diagram that is responsible for the sign change of the double bubble observed in the 2PPI effective action as written in eq.~\eqref{eq:G2PPI}.

At the two-loop level, the bi-local source takes the form
\begin{equation}
\label{eq:twoloopK2PPI}
\mathcal{K}_{xy}[\phi,\Delta]\ =\ \bigg[-\,2\hbar\,\frac{\delta \Gamma_{1}[\varphi,\mathcal{G}]}{\delta \varphi_x^2}\:-\:2\hbar^2\,\frac{\delta \Gamma_{2}[\varphi,\mathcal{G}]}{\delta \varphi_x^2}\bigg]\delta^{(4)}_{xy}\;.
\end{equation}
As we will see, two-loop diagrams will arise from both $\Gamma_1[\varphi,\mathcal{G}]$ and $\Gamma_2[\varphi,\mathcal{G}]$. The contribution to eq.~\eqref{eq:twoloopK2PPI} from $\Gamma_1[\varphi,\mathcal{G}]$ (expanding further) gives
\begin{equation}
\label{eq:VCGam1}
-\,2\hbar\,\frac{\delta \Gamma_{1}[\varphi,\mathcal{G}]}{\delta \varphi_x^2} \ =\ \hspace{0.5em}\raisebox{-0.45cm}{\includegraphics[scale=0.75]{self1.pdf}}\hspace{0.5em}-\:\hbar\,\frac{\delta}{\delta\varphi^2_x}\,\mathcal{K}_{yz}[\phi,\Delta]\mathcal{G}_{zy}+\:\hbar\,\mathcal{G}_{yz}\,\frac{\delta\mathcal{K}_{zy}[\phi,\Delta]}{\delta\varphi^2_x}\;,
\end{equation}
where the second term originates from the additional double-bubble diagram in eq.~\eqref{eq:Gamma1bub} and the third term from the implicit dependence of the propagator $\mathcal{G}$, running in the loops, on $\mathcal{K}_{xy}[\phi,\Delta]$.
The contribution to eq.~\eqref{eq:twoloopK2PPI} from $\Gamma_2[\varphi,\mathcal{G}]$ gives rise to the two-loop diagrams
\begin{equation}
\label{eq:VCGam2}
-\,2\hbar^2\,\frac{\delta \Gamma_{2}[\varphi,\mathcal{G}]}{\delta \varphi_x^2} \ =\ \hspace{0.5em}\raisebox{-2em}{\includegraphics[scale=0.75]{tad2.pdf}}\hspace{0.5em}+\hspace{0.5em}\raisebox{-0.42cm}{\includegraphics[scale=0.75]{tad3.pdf}}\hspace{0.5em}+\:\mathcal{O}(\hbar^3)\;,
\end{equation}
where the $\mathcal{O}(\hbar^3)$ terms arise in analogy to the $\mathcal{O}(\hbar^2)$ terms in eq.~\eqref{eq:VCGam1} through the implicit dependence on $\mathcal{K}_{xy}[\phi,\Delta]$, as well as higher-order loops, which we do not consider explicitly.

The presence of the double-bubble tadpole in eq.~\eqref{eq:VCGam2} would look to be a serious problem, since this diagram is already counted in the first diagram on the right-hand side of eq.~\eqref{eq:VCGam1} by virtue of the resummation of the one-loop tadpole insertions. However, by successive substitution of $\mathcal{K}_{xy}[\phi,\Delta]$ from eq.~\eqref{eq:VCGam1} back into itself, we find
\begin{equation}
-\,2\hbar\,\frac{\delta \Gamma_{1}[\varphi,\mathcal{G}]}{\delta \varphi_x^2} \ =\ \raisebox{-0.45cm}{\includegraphics[scale=0.75]{self1.pdf}}\hspace{0.5em}-\:\hspace{0.5em}\raisebox{-2em}{\includegraphics[scale=0.75]{tad2.pdf}}\hspace{0.5em}+\:\mathcal{O}(\hbar^3)\;.
\end{equation}
Thus, the problematic double-bubble tadpole diagram cancels between eqs.~\eqref{eq:VCGam1} and~\eqref{eq:VCGam2}, giving the two-loop bi-local source
\begin{equation}
\label{eq:2loop2PPIK}
\mathcal{K}_{xy}[\phi,\Delta]\ =\ \raisebox{-0.45cm}{\includegraphics[scale=0.75]{self1.pdf}}\hspace{0.5em}+\hspace{0.5em}\raisebox{-0.42cm}{\includegraphics[scale=0.75]{tad3.pdf}}\hspace{0.5em}\;.
\end{equation}
Comparing with eq.~\eqref{eq:Dxx}, we see that eq.~\eqref{eq:2loop2PPIK} precisely matches the two-loop 2PPI result, since $\mathcal{K}_{xy}[\phi,\Delta]$ plays the role of $-\,\hbar\,\lambda\, \Delta_{xx} /2$ in the 2PPI Schwinger-Dyson equation.

Before concluding this section, we consider the form of the local source $\mathcal{J}_x[\phi,\Delta]$. As was the case at one-loop, the external source must satisfy the consistency relation [cf.~eq.~\eqref{eq:deltaS}], constraining
\begin{equation}
\label{eq:2PPIconsist}
\mathcal{J}_x[\phi,\Delta]\:+\:\mathcal{K}_{xy}[\phi,\Delta]\varphi_y\ =\ -\,\hbar\,\frac{\delta \Gamma_1[\phi,\Delta]}{\delta \phi_x}\bigg|_{\varphi,\mathcal{G}}\:-\:\hbar^2\,\frac{\delta\Gamma_2[\phi,\Delta]}{\delta \phi_x}\bigg|_{\varphi,\mathcal{G}}
\end{equation}
The first term on the right-hand side of eq.~\eqref{eq:2PPIconsist} gives
\begin{equation}
\label{eq:2loop1}
-\,\hbar\,\frac{\delta \Gamma_1[\phi,\Delta]}{\delta \phi_x}\bigg|_{\varphi,\mathcal{G}}\ =\ \raisebox{-0.45cm}{\includegraphics[scale=0.75]{tad.pdf}}\hspace{0.5em}\;,
\end{equation}
and the second term gives
\begin{equation}
\label{eq:2loop2}
-\,\hbar^2\,\frac{\delta \Gamma_{2}[\phi,\Delta]}{\delta \phi_x}\bigg|_{\varphi,\mathcal{G}} \ =\ \hspace{0.5em}\raisebox{-0.36cm}{\includegraphics[scale=0.75]{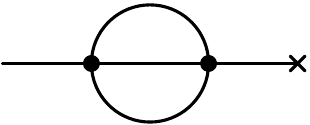}}\hspace{0.5em}\;.
\end{equation}
Inserting the diagrammatic expressions in eqs.~\eqref{eq:2loop1} and~\eqref{eq:2loop2} back into eq.~\eqref{eq:2PPIconsist}, we obtain
\begin{equation}
\label{eq:JplusK2PPI}
\mathcal{J}_x[\phi,\Delta]\:+\:\mathcal{K}_{xy}[\phi,\Delta]\varphi_y\ =\ \raisebox{-0.45cm}{\includegraphics[scale=0.75]{tad.pdf}}\hspace{0.5em}+\hspace{0.5em}\raisebox{-0.36cm}{\includegraphics[scale=0.75]{tad4x.pdf}}\hspace{0.5em}.
\end{equation}
Hence, given the form of the bi-local source in eq.~\eqref{eq:2loop2PPIK}, we find
\begin{equation}
\mathcal{J}_x[\phi,\Delta]\ = \hspace{0.5em}\raisebox{-0.35cm}{\includegraphics[scale=0.75]{tad4x.pdf}}\hspace{0.5em}-\hspace{0.5em}\raisebox{-0.42cm}{\includegraphics[scale=0.75]{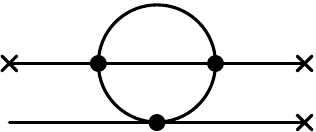}}\hspace{0.5em}.
\end{equation}
By virtue of eqs.~\eqref{eq:deltaS} and~\eqref{eq:JplusK2PPI}, we see that we recover the usual quantum equation of motion for the physical one-point function.

\section{Global symmetries}
\label{sec:syms}

In this final example, we will illustrate how this external-source method can be used to preserve symmetries in truncations of the effective action. Specifically, we consider the following globally $O(N)$ symmetric model:
\begin{equation}
\mathcal{L}_x\ =\ \frac{1}{2!}\,\big(\partial_{\mu}\Phi^i_x\big)^2\:+\:\frac{1}{2!}\,m^2\,\big(\Phi^i_x\big)^2\:+\:\frac{\lambda}{(2!)^2}\,\big(\Phi^i_x\big)^2\big(\Phi^j_x\big)^2\;,
\end{equation}
in which $\Phi^i_x=(\Phi^1_x,\Phi^2_x,\dots,\Phi^N_x)$ is the $O(N)$ scalar multiplet and repeated indices are summed over. Taking $m^2<0$, this $O(N)$ symmetry is spontaneously broken, and we obtain one massive mode (the ``Higgs'') and $N-1$ massless Goldstone modes~\cite{Englert:1964et,Higgs:1964pj,Guralnik:1964eu}.

For this model, the 2PI effective action is defined by the Legendre transform
\begin{equation}
\Gamma[\phi,\Delta]\ =\ \mathrm{max}_{J,\:K}\bigg[-\,\hbar\,\ln Z[J,K]\:+\:J_x^i\,\phi^i_x\:+\:\frac{1}{2}\,K_{xy}^{ij}\,\Big(\phi^i_x\,\phi^j_y\:+\:\hbar\,\Delta^{ij}_{xy}\Big)\bigg]\;,
\end{equation}
where we have introduced the multiplet of local sources $J_x^i$ and the matrix of bi-local sources $K^{ij}_{xy}$. After performing the extremization with respect to $J^i_x$ and $K^{ij}_{xy}$, the 2PI effective action takes the following form:
\begin{equation}
\label{eq:2PIO2}
\Gamma[\phi,\Delta]\ =\ -\,\hbar\,\ln Z[\mathcal{J},\mathcal{K}]\:+\:\mathcal{J}_x^i[\phi,\Delta]\,\phi^i_x\:+\:\frac{1}{2}\,\mathcal{K}_{xy}^{ij}[\phi,\Delta]\,\Big(\phi^i_x\,\phi^j_y\:+\:\hbar\,\Delta^{ij}_{xy}\Big)\;,
\end{equation}
where the various one- and two-point functions are given by [cf.~eq.~\eqref{eq:onetwodef}]
\begin{subequations}
\begin{align}
\phi^i_x\ &=\ \hbar\,\frac{\delta \ln Z[\mathcal{J},\mathcal{K}]}{\delta J_x^i}\bigg|_{J\,=\,\mathcal{J},\:K\,=\,\mathcal{K}}\;,\\
\Delta^{ij}_{xy}\ &=\ 2\,\frac{\delta \ln Z[\mathcal{J},\mathcal{K}]}{\delta K^{ij}_{xy}}\bigg|_{J\,=\,\mathcal{J},\:K\,=\,\mathcal{K}}\:-\:\frac{1}{\hbar}\,\phi^i_x\,\phi^j_y\;.
\end{align}
\end{subequations}
The external sources $\mathcal{J}_x^i[\phi,\Delta]$ and $\mathcal{K}_{xy}^{ij}[\phi,\Delta]$ are determined by the stationarity conditions
\begin{subequations}
\label{eq:statON}
\begin{align}
\frac{\delta \Gamma[\phi,\Delta]}{\delta \phi^i_x}\ &=\ \mathcal{J}_x^i[\phi,\Delta]\:+\:\mathcal{K}_{xy}^{ij}[\phi,\Delta]\,\phi^j_y\;,\\
\frac{\delta \Gamma[\phi,\Delta]}{\delta \Delta^{ij}_{xy}}\ &=\ \frac{\hbar}{2}\,\mathcal{K}_{xy}^{ij}[\phi,\Delta]\;,
\end{align}
\end{subequations}
analogous to eq.~\eqref{eq:diffs}.

From the symmetry of eq.~\eqref{eq:2PIO2} under $O(N)$ transformations, we can derive the 2PI Ward identity, which is given by
\begin{equation}
\frac{\delta \Gamma[\phi,\Delta]}{\delta \phi_x^i}\,T_{ij}^a\,\phi^j_x\:+\:\frac{\delta \Gamma[\phi,\Delta]}{\delta\Delta^{ij}_{xy}}\,\Big(T_{ik}^a\,\Delta^{kj}_{xy}\:+\:T_{jl}^a\,\Delta_{xy}^{il}\Big)\ =\ 0\;,
\end{equation} 
where $T_{ij}^a$ are the generators of the $O(N)$ group. By virtue of the stationarity conditions in eq.~\eqref{eq:statON}, the 2PI Ward identity can be re-expressed in terms of the external bi-local sources $\mathcal{K}^{ij}_{xy}[\phi,\Delta]$ as
\begin{equation}
\label{eq:Ward}
\frac{\delta \Gamma[\phi,\Delta]}{\delta \phi_x^i}\,T_{ij}^a\,\phi^j_x\:+\:\frac{\hbar}{2}\,\mathcal{K}_{xy}^{ij}[\phi,\Delta]\,\Big(T_{ik}^a\,\Delta^{kj}_{xy}\:+\:T_{jl}^a\,\Delta_{xy}^{il}\Big)\ =\ 0\;.
\end{equation}
In what follows, we will consider the case $N=2$ for definiteness and simplicity.

\paragraph{\bfseries Symmetric gauge} It will prove illustrative to consider first the following manifestly symmetric gauge choice:
\begin{equation}
\label{eq:symmgauge}
\Phi_x\ =\ \begin{pmatrix} \varphi^H/\sqrt{2} + \hbar^{1/2}\hat{\Phi}^1_x \\ \varphi^H/\sqrt{2} + \hbar^{1/2}\hat{\Phi}^2_x\end{pmatrix}\;.
\end{equation}
The notation for the physical one-point functions $\varphi^1=\varphi^2=\varphi^{H}/\sqrt{2}$ has been chosen for later convenience. We continue in complete analogy to section~\ref{sec:ext}, eliminating $\phi^i_x$ and $\Delta^{ij}_{xy}$ in favour of $\varphi^i_x$ and $\mathcal{G}_{xy}^{ij}$. The only modification to the 2PI effective action in eq.~\eqref{eq:Gammaexp} is the presence of the additional field-space structure, e.g.
\begin{equation}
\Gamma_1[\varphi,\mathcal{G}]\ =\ \frac{1}{2}\,\mathrm{tr}\,\big[\ln\,\mathrm{det}_{ij}\,\bm{\mathcal{G}}^{-1}\ast \bm{G_0}\:+\:\bm{G}^{-1}\ast\bm{\mathcal{G}}\:-\:\bm{1} \big]\;,
\end{equation}
where we have used boldface symbols for $2\times2$ matrices in field space and $\mathrm{det}_{ij}$ for the determinant in field space. 

In order to obtain the quantum equations of motion, the local and bi-local sources are constrained as in section~\ref{sec:CJT} to be the extremal configurations [see~eq.~\eqref{eq:CJTeoms}]. Thus, by virtue of the consistency relation in eq.~\eqref{eq:Jconsist}, we require that
\begin{subequations}
\begin{gather}
\mathcal{J}^{i}_x[\phi,\Delta]\ =\ -\:\hbar\,\frac{\delta \Gamma_1[\phi,\Delta]}{\delta \phi^i_x}\bigg|_{\varphi,\,\mathcal{G}}\:+\:2\,\hbar\,\frac{\delta \Gamma_2[\phi,\Delta]}{\delta \Delta^{ij}_{xy}}\bigg|_{\varphi,\,\mathcal{G}}\varphi^j_y\;,\\
\mathcal{K}^{ij}_{xy}[\phi,\Delta]\ =\ -\:2\,\hbar\,\frac{\delta \Gamma_2[\phi,\Delta]}{\delta \Delta^{ij}_{xy}}\bigg|_{\varphi,\,\mathcal{G}}\;,
\end{gather}
\end{subequations}
generalizing eqs.~\eqref{eq:KCJT} and~\eqref{eq:JCJT}. Together, these yield the equation of motion for the background field configuration
\begin{equation}
\label{eq:Higgseom}
-\:\partial_x^2\varphi^{H}_x\:+\:m^2\varphi^H_x\:+\:\lambda(\varphi^H)^3\:-\:\sqrt{2}\,\mathcal{J}_x^1[\phi,\Delta]\:-\:\Big(\mathcal{K}_{xy}^{11}[\phi,\Delta]\:+\:\mathcal{K}_{xy}^{12}[\phi,\Delta]\Big)\varphi^H_y\ =\ 0
\end{equation}
and the physical inverse two-point functions
\begin{equation}
\mathcal{G}^{-1,\,ij}_{xy}[\phi,\Delta]\ =\ G^{-1,\,ij}_{xy}(\varphi)\:-\:\mathcal{K}^{ij}_{xy}[\phi,\Delta]\;,
\end{equation}
where
\begin{subequations}
\begin{gather}
G^{-1,\,11}_{xy}(\varphi)\ =\ G^{-1,\,22}_{xy}(\varphi)\ =\ \delta^{(4)}_{xy}\Big[-\,\partial_x^2\:+\:m^2\:+\:2\,\lambda\,\big(\varphi^H_x\big)^2\Big]\;,\\
G^{-1,\,12}_{xy}(\varphi)\ =\ G^{-1,\,21}_{xy}(\varphi)\ =\ \delta^{(4)}_{xy}\,\lambda\,\big(\varphi^H_x\big)^2\;.
\end{gather}
\end{subequations}
Assuming a constant background field $\varphi^H_x\equiv \varphi^H$, we may solve explicitly for the tree-level propagators in momentum space:
\begin{subequations}
\label{eq:symprops}
\begin{align}
G^{11}_{k}\ =\ G^{22}_{k}\ =\  \frac{k^2+m^2+2\lambda\big(\varphi^H\big)^2}{\big[k^2+m^2+2\lambda\big(\varphi^H\big)^2\big]^2-\big[\lambda\big(\varphi^H\big)^2\big]^2}\;,\\
G^{12}_{k}\ =\ G^{21}_k\ = \ \frac{-\lambda\big(\varphi^H\big)^2}{\big[k^2+m^2+2\lambda\big(\varphi^H\big)^2\big]^2-\big[\lambda\big(\varphi^H\big)^2\big]^2}\;.
\end{align}
\end{subequations}

We now consider the structure of the 2PI Ward identity. For $N=2$, we have only one generator: $T^1=\sigma_2$, where $\sigma_2$ is the second of the Pauli matrices. Expanding the first term in the 2PI Ward identity [eq.~\eqref{eq:Ward}] about the physical one- and two-point functions $\varphi^i$ and $\mathcal{G}^{ij}$ to order $\hbar^2$, we may show that it vanishes by virtue of the extremal conditions and the anti-symmetry of $\sigma_2$. Thus, we are left with the following constraint on the bi-local sources
\begin{equation}
\label{eq:WIcons}
\mathcal{K}_{xy}^{ij}[\phi,\Delta]\,\Big(T_{ik}^a\,\Delta^{kj}_{xy}\:+\:T_{jl}^a\,\Delta_{xy}^{il}\Big)\ =\ 0\;,
\end{equation}
having the explicit form
\begin{equation}
\label{eq:wardpart}
\Big(\mathcal{K}_{xy}^{11}[\phi,\Delta]\:-\:\mathcal{K}_{xy}^{22}[\phi,\Delta]\Big)\,\Delta^{12}_{xy}\:-\:\,\mathcal{K}^{12}_{xy}[\phi,\Delta]\,\Big(\Delta^{11}_{xy}\:-\:\Delta^{22}_{xy}\Big)\ =\ 0\;.
\end{equation}
Herein, we have used the fact that $\Delta^{ij}_{xy}=\Delta^{ji}_{yx}$ and $\mathcal{K}^{ij}_{xy}=\mathcal{K}^{ji}_{yx}$. By virtue of the manifest $O(2)$ symmetry of this gauge, we have
\begin{equation}
\mathcal{K}^{11}_{xy}[\phi,\Delta]\ =\ \mathcal{K}^{22}_{xy}[\phi,\Delta]\;,
\end{equation}
and, from the stationarity conditions,
\begin{equation}
\Delta^{11}_{xy}\ =\ \Delta^{22}_{xy}\;.
\end{equation}
Thus, we see that eq.~\eqref{eq:wardpart} and thereby the 2PI Ward identity [eq.~\eqref{eq:Ward}] are immediately satisfied.

\paragraph{\bfseries Unitary gauge} We now consider the same expressions in the unitary gauge~\cite{Weinberg:1971fb,Weinberg:1973ew}. We can rotate to this gauge by means of the orthogonal transformation
\begin{equation}
\Phi_x\ \longrightarrow\ \Phi'_x\ =\ \frac{1}{\sqrt{2}}\begin{pmatrix} 1 & -\,1 \\ 1 & 1\end{pmatrix}\Phi_x\ =\ \begin{pmatrix} \hbar^{1/2}\hat{\Phi}^G_x \\ \varphi^H + \hbar^{1/2}\hat{\Phi}^H_x \end{pmatrix}\;,
\end{equation}
where
\begin{equation}
\hat{\Phi}^{H(G)}_x\ =\ \frac{1}{\sqrt{2}}\,\Big(\hat{\Phi}^1_x\:+(-)\:\hat{\Phi}^2_x\Big)\;.
\end{equation}
Here, the superscripts $H$ and $G$ indicate the Higgs and Goldstone modes. In the case of a constant background field configuration $\varphi^H_x=\varphi^H$, we now have the inverse of the tree-level momentum-space propagators
\begin{subequations}
\begin{gather}
G^{-1,\,HH}_{k}\ = \ G^{-1,\,11}_k\:+\:G^{-1,\,12}_k\ =\ k^2\:+\:m^2\:+\:3\,\lambda\,\big(\varphi^H_x\big)^2\;,\\
\label{eq:GGG}
G^{-1,\,GG}_{k}\ =\ G^{-1,\,11}_k\:-\:G^{-1,\,12}_k\ =\ k^2\:+\:m^2\:+\:\lambda\,\big(\varphi^H_x\big)^2\;,\\
G^{-1,\,HG}_{k}\  =\ G^{-1,\,GH}_{k}\ =\ 0\;.
\end{gather}
\end{subequations}
Since the vacuum expectation value is given by $\varphi^H=v=\pm\,|m|/\sqrt{\lambda}$ at tree-level, we see that the Goldstone propagator is massless, as we would expect.

The equation of motion for the background field is given by
\begin{equation}
\label{eq:Hcons}
\frac{\delta \Gamma[\phi,\Delta]}{\delta \phi^H_x}\bigg|_{\varphi,\mathcal{G}}\ =\ \frac{\delta S[\varphi]}{\delta \varphi^H_x}\:-\:\mathcal{J}_x^H[\phi,\Delta]\:-\:\mathcal{K}_{xy}^{HH}[\phi,\Delta]\,\varphi^H_y\ =\ 0\;.
\end{equation}
Comparing this with eq.~\eqref{eq:Higgseom}, we see that
\begin{equation}
\label{eq:KHH}
\mathcal{K}^{HH}_{xy}[\phi,\Delta]\ =\ \mathcal{K}_{xy}^{11}[\phi,\Delta]\:+\:\mathcal{K}_{xy}^{12}[\phi,\Delta]\ =\ -\:2\,\hbar\,\frac{\delta \Gamma_2[\phi,\Delta]}{\delta \Delta^{HH}_{xy}}\bigg|_{\varphi,\mathcal{G}}\;.
\end{equation}
Thus, diagrammatically, we have
\begin{equation}
\mathcal{K}^{HH}_{xy}[\phi,\Delta]\ = \hspace{0.5em}\raisebox{-0.45cm}{\includegraphics[scale=0.75]{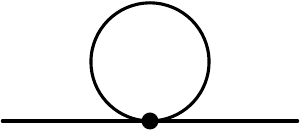}}\hspace{0.5em}\:+\:\hspace{0.5em}\raisebox{-0.45cm}{\includegraphics[scale=0.75]{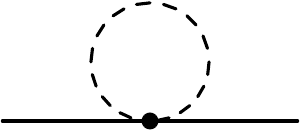}}\hspace{0.5em}+\hspace{0.5em}\raisebox{-0.52cm}{\includegraphics[scale=0.75]{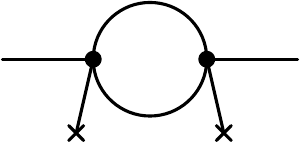}}\hspace{0.5em}\:+\:\hspace{0.5em}\raisebox{-0.52cm}{\includegraphics[scale=0.75]{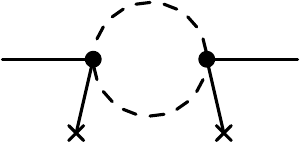}}\hspace{0.5em}\;,
\end{equation}
where solid lines now correspond to Higgs propagators and dashed lines to Goldstone propagators. In addition, the vanishing of the Goldstone boson component $\varphi^G_x$ implies the constraint
\begin{equation}
\label{eq:Gsources}
\mathcal{J}_x^G[\phi,\Delta]\:+\:\mathcal{K}^{GH}_{xy}[\phi,\Delta]\,\varphi_y^H\ =\ 0\;.
\end{equation}
Having imposed that $\mathcal{G}^{HG}_{xy}$ is the extremal configuration, we immediately find that
\begin{equation}
\mathcal{K}_{xy}^{HG}[\phi,\Delta]\ =\ -\:2\,\hbar\,\frac{\delta \Gamma_2[\phi,\Delta]}{\delta \Delta^{HG}_{xy}}\bigg|_{\varphi,\mathcal{G}}\ =\ 0
\end{equation}
and, equivalently, $\mathcal{K}^{GH}_{xy}[\phi,\Delta]\ =\ 0$.
It then follows from eq.~\eqref{eq:Gsources} that
\begin{equation}
\mathcal{J}^G_x[\phi,\Delta]\ =\ 0\;.
\end{equation}
The consistency relation in eq.~\eqref{eq:Hcons} [cf.~eq.~\eqref{eq:Jconsist}] fixes
\begin{equation}
\mathcal{J}_x^H[\phi,\Delta]\ =\ -\:\hbar\,\frac{\delta \Gamma_1[\phi,\Delta]}{\delta \phi^H_x}\bigg|_{\varphi,\mathcal{G}}\:-\:\mathcal{K}_{xy}^{HH}[\phi,\Delta]\,\varphi^H_y\;.
\end{equation}
We are then left with only \emph{one} source to fix: $\mathcal{K}^{GG}_{xy}[\phi,\Delta]$. In what follows, we will show that by using the Ward identity to constrain the form of this source, rather than associating it with the 2PI Goldstone self-energy, we may preserve the symmetry properties of truncations of the effective action. It is this use of the Ward identity to directly constrain the truncation of the effective action that is in the spirit of the PT symmetry-improved effective action~\cite{Pilaftsis:2013xna, Pilaftsis:2015cka,Pilaftsis:2015bbs}, and we will see that we obtain comparable results.

In order to understand the constraint on $\mathcal{K}^{GG}_{xy}[\phi,\Delta]$ from the 2PI Ward identity, we rotate the symmetric gauge choice in eq.~\eqref{eq:symmgauge} infinitesimally close to the unitary gauge via the transformation
\begin{equation}
\Phi_x\ \longrightarrow\ \Phi'_x\ =\ \frac{1}{\sqrt{2}}\begin{pmatrix} 1 & -1+\epsilon \\ 1-\epsilon & 1\end{pmatrix}\Phi_x\;,
\end{equation}
where $\epsilon\to 0^+$. In this gauge, neglecting terms $\mathcal{O}(\epsilon^2)$, eq.~\eqref{eq:wardpart} reads
\begin{equation}
\label{eq:epsconst}
\Big(\mathcal{K}_{xy}^{HH}[\phi,\Delta]\:-\:\mathcal{K}_{xy}^{GG}[\phi,\Delta]\Big)\,\Delta^{GH(\epsilon)}_{xy}\:-\:\,\mathcal{K}^{GH(\epsilon)}_{xy}[\phi,\Delta]\,\Big(\Delta^{HH}_{xy}\:-\:\Delta^{GG}_{xy}\Big)\ =\ 0\;,
\end{equation}
where
\begin{equation}
\Delta^{GH(\epsilon)}_{xy}\ =\ \epsilon\,\Delta^{12}_{xy}\;,\qquad \mathcal{K}^{GH(\epsilon)}_{xy}\ =\ \epsilon\, \mathcal{K}^{12}_{xy}\;.
\end{equation}
Here, the Arabic numerals refer to the field modes of the symmetric gauge. Thus, from eq.~\eqref{eq:epsconst}, we find the constraint
\begin{equation}
\label{eq:witheps}
\Big(\mathcal{K}_{xy}^{HH}[\phi,\Delta]\:-\:\mathcal{K}_{xy}^{GG}[\phi,\Delta]\:-\:2\,\mathcal{K}^{12}_{xy}[\phi,\Delta]\Big)\,\Delta^{12}_{xy}\ =\ 0\;,
\end{equation}
which holds only if
\begin{equation}
\label{eq:Keq}
\mathcal{K}^{12}_{xy}[\phi,\Delta]\ =\ \frac{1}{2}\Big(
\mathcal{K}_{xy}^{HH}[\phi,\Delta] - \mathcal{K}_{xy}^{GG}[\phi,\Delta]\Big)\;.
\end{equation}
This is, of course, as we would expect, if the bi-local source transforms as a rank-$2$ tensor of $O(2)$, and it would appear that we have gained very little. Nevertheless, we will continue and consider the explicit form of
\begin{equation}
\mathcal{K}^{12}_{xy}[\phi,\Delta]\ =\ -\,2\hbar\,\frac{\delta\Gamma_2[\phi,\Delta]}{\delta\Delta^{12}_{xy}}\bigg|_{\varphi,\mathcal{G}}
\end{equation}
at the lowest order in perturbation theory, i.e.~at one-loop with tree-level propagators.

To this end, we isolate the tadpole and bubble contributions to $\mathcal{K}^{12}_{xy}[\phi,\Delta]$, writing
\begin{equation}
\mathcal{K}^{12}_{xy}[\phi,\Delta]\ \equiv\ \mathcal{K}^{12\,(\mathrm{tad})}_{xy}[\phi,\Delta]\:+\:\mathcal{K}^{12\,(\mathrm{bub})}_{xy}[\phi,\Delta]\;,
\end{equation}
where
\begin{subequations}
\begin{align}
\label{eq:tad12diag}
\mathcal{K}^{12\,(\mathrm{tad})}_{xy}[\phi,\Delta]\ &=\ \frac{1}{2}\,\Bigg[ \hspace{0.5em}\raisebox{-0.45cm}{\includegraphics[scale=0.75]{HHtadHH.pdf}}\hspace{0.5em}\:+\:\hspace{0.5em}\raisebox{-0.45cm}{\includegraphics[scale=0.75]{HHtadGG.pdf}}\hspace{0.5em}\nonumber\\&\hspace{7em} -\:\hspace{0.5em}\raisebox{-0.45cm}{\includegraphics[scale=0.75]{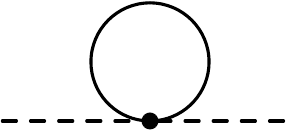}}\hspace{0.5em}\:-\:\hspace{0.5em}\raisebox{-0.45cm}{\includegraphics[scale=0.75]{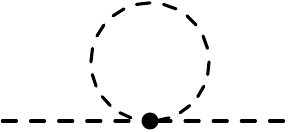}}\hspace{0.5em}\Bigg]\;,\\[1em]
\mathcal{K}^{12\,(\mathrm{bub})}_{xy}[\phi,\Delta]\ &=\ \frac{1}{2}\,\Bigg[ \hspace{0.5em}\raisebox{-0.52cm}{\includegraphics[scale=0.75]{HH.pdf}}\hspace{0.5em}+\hspace{0.5em}\raisebox{-0.52cm}{\includegraphics[scale=0.75]{GG.pdf}}\hspace{0.5em}-\hspace{0.5em}\raisebox{-0.52cm}{\includegraphics[scale=0.75]{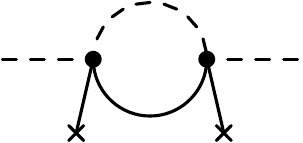}}\hspace{0.5em}\Bigg]\;.
\end{align}
\end{subequations}
The solid lines now correspond to \emph{tree-level} Higgs propagators and dashed lines to \emph{tree-level} Goldstone propagators. The tadpole contribution to $\mathcal{K}^{12}_{xy}[\phi,\Delta]$ may be written explicitly as
\begin{align}
\label{eq:K12tad}
\mathcal{K}_{xy}^{12\,(\mathrm{tad})}[\phi,\Delta]\ &=\ -\,2\,\hbar\,\lambda\,\mathcal{G}_{xx}^{12}\,\delta^{(4)}_{xy}\nonumber\\ &=\ 2\,\hbar\,\lambda^2(\varphi^H)^2\, \delta^{(4)}_{xy} \int_k\frac{1}{k^2+m^2+3\lambda(\varphi^H)^2}\,\frac{1}{k^2+m^2+\lambda(\varphi^H)^2}\;.
\end{align}
By inspection, we see that this is just the bubble contribution to the Goldstone self-energy with the external momentum set to zero, i.e.~its local part. It is this observation that will allow us to re-organize the Hartree-Fock approximation, ensuring that the 2PI Ward identity is satisfied and the Goldstone boson remains massless in the SSB phase.

Having observed that the tadpole diagrams in eq.~\eqref{eq:tad12diag} may be reinterpreted as a single bubble diagram (with vanishing external momentum), we may write the full form of $\mathcal{K}^{12}_{xy}[\phi,\Delta]$ as
\begin{align}
\label{eq:K12}
\mathcal{K}_{xy}^{12}[\phi,\Delta]\ &=\ \frac{1}{2}\,\Bigg[ \hspace{0.5em}\raisebox{-0.52cm}{\includegraphics[scale=0.75]{HH.pdf}}\hspace{0.5em}\:+\:\hspace{0.5em}\raisebox{-0.52cm}{\includegraphics[scale=0.75]{GG.pdf}}\hspace{0.5em}\nonumber\\&\hspace{7em}-\:\hspace{0.5em}\raisebox{-0.52cm}{\includegraphics[scale=0.75]{GH.pdf}}\hspace{0.5em}\:+\:\hspace{0.5em}\raisebox{-0.52cm}{\includegraphics[scale=0.75]{GH.pdf}}\hspace{0.5em}\Bigg|_{\mathrm{loc}}\:\Bigg]\;.
\end{align}
Rearranging eq.~\eqref{eq:Keq}, we have
\begin{equation}
\mathcal{K}^{GG}_{xy}[\phi,\Delta]\ =\ \mathcal{K}^{HH}_{xy}[\phi,\Delta]\:-\:2\mathcal{K}^{12}_{xy}[\phi,\Delta]\;.
\end{equation}
Subsequently, using eq.~\eqref{eq:KHH}, we see that the bubble contributions to the Higgs self-energy are cancelled by the first two diagrams in eq.~\eqref{eq:K12}. Classing the remaining diagrams in eq.~\eqref{eq:K12} as the \emph{bubble} contribution to the Goldstone self-energy $\Pi^{GG({\rm bub})}_{xy}$, we have
\begin{align}
&-\,\frac{\hbar}{2}\,\Pi^{GG\,(\mathrm{bub})}_{xy}\ \equiv\ \frac{1}{2}\,\Bigg[\hspace{0.5em}\raisebox{-0.52cm}{\includegraphics[scale=0.75]{GH.pdf}}\hspace{0.5em}\:-\:\hspace{0.5em}\raisebox{-0.52cm}{\includegraphics[scale=0.75]{GH.pdf}}\hspace{0.5em}\Bigg|_{\mathrm{loc}}\:\Bigg]\nonumber\\[1em]&= \ 2\,\hbar\,\lambda^2\,(\varphi^H)^2\int_k\frac{2\,k\cdot p-p^2}{\big(k^2+m^2+\lambda(\varphi^H)^2\big)\big(k^2+m^2+3\lambda(\varphi^H)^2\big)\big((k-p)^2+m^2+3\lambda(\varphi^H)^2\big)}\;.
\end{align}
Wick rotating to Minkowski space, this integral is proportional to $p^2$ by Lorentz invariance and therefore vanishes on-shell ($p^2=0$). As a result, we see that $\mathcal{K}^{12}_{xy}[\phi,\Delta]$ does not contain any contribution to a would-be Goldstone mass.

Having reclassified the tadpole contribution to $\mathcal{K}^{12}_{xy}[\phi,\Delta]$ [see~eqs.~\eqref{eq:tad12diag} and~\eqref{eq:K12tad}] as a \emph{bubble} contribution to the Goldstone self-energy, we are left with the following constraint on the remaining \emph{tadpole} contribution to $\mathcal{K}^{GG}_{xy}[\phi,\Delta]$ from eq.~\eqref{eq:witheps}:
\begin{equation}
\label{eq:Ktadconst}
\mathcal{K}^{GG\,(\mathrm{tad})}_{xy}[\phi,\Delta]\ =\ \mathcal{K}^{HH\,(\mathrm{tad})}_{xy}[\phi,\Delta]\;.
\end{equation}
As we will see next, this identity is sufficient to satisfy the 2PI Ward identity and obtain a massless Goldstone boson algebraically at one-loop order in the Hartree-Fock approximation.

\paragraph{\bfseries Hartree-Fock approximation} The HF approximation~\cite{Hartree,Fock} corresponds to keeping only the double-bubble diagram in the 2PI effective action. At the level of the self-energy, this amounts to keeping only the tadpole contributions.

Imposing the constraint in eq.~\eqref{eq:Ktadconst}, the equations of motion for the physical one- and two-point functions are as follows:
\begin{subequations}
\begin{gather}
\frac{\delta \Gamma[\phi,\Delta]}{\delta \phi^{H(G)}_x}\bigg|_{\varphi,\mathcal{G}}\ =\ 0\;,\\
\frac{\delta \Gamma[\phi,\Delta]}{\delta \Delta^{HH(HG)}_{xy}}\bigg|_{\varphi,\mathcal{G}}\ =\ 0\;,\\
\label{eq:Goldstoneeom}
\frac{\delta \Gamma[\phi,\Delta]}{\delta \Delta_{xy}^{GG}}\bigg|_{\varphi,\mathcal{G}}\ =\ \hbar^2\,\bigg[\frac{\delta \Gamma^{(\mathrm{HF})}_2[\phi,\Delta]}{\delta \Delta_{xy}^{GG}}\:-\:\frac{\delta \Gamma^{(\mathrm{HF})}_2[\phi,\Delta]}{\delta \Delta_{xy}^{HH}}\bigg]_{\varphi,\mathcal{G}}\;,
\end{gather}
\end{subequations}
where $\Gamma_2^{(\mathrm{HF})}[\phi,\Delta]$ is the double-bubble contribution to $\Gamma_2[\phi,\Delta]$.
Note that the equation of motion for the Goldstone-boson two-point function $\mathcal{G}^{GG}_{xy}$ does \emph{not} correspond to the standard Schwinger-Dyson equation, since the right-hand side of eq.~\eqref{eq:Goldstoneeom} is non-zero in the SSB phase. Therefore, the physical limit, consistent with Goldstone's theorem in the SSB phase for the HF approximation, is obtained for a \emph{non-vanishing} bi-local source
\begin{equation}
\label{eq:GGsource}
\mathcal{K}^{GG}_{xy}[\varphi,\mathcal{G}]\ =\ 2\,\hbar\,\bigg[\frac{\delta \Gamma^{(\mathrm{HF})}_2[\phi,\Delta]}{\delta \Delta_{xy}^{GG}}\:-\:\frac{\delta \Gamma^{(\mathrm{HF})}_2[\phi,\Delta]}{\delta \Delta_{xy}^{HH}}\bigg]_{\varphi,\mathcal{G}}\;.
\end{equation}

We may verify that eq.~\eqref{eq:GGsource} is entirely consistent with eq.~\eqref{eq:Ktadconst}. In order to do so, we expand the left-hand side of
\begin{equation}
\mathcal{K}^{GG}_{xy}[\phi,\Delta]\ =\ -\:2\,\hbar\,\frac{\delta \Gamma_2^{(\mathrm{HF})}[\phi,\Delta]}{\delta\Delta^{HH}_{xy}}\bigg|_{\varphi,\mathcal{G}}
\end{equation}
around $\varphi^i_x$ and $\mathcal{G}^{ij}_{xy}$, in analogy to eq.~\eqref{eq:Kexp}. At order $\hbar$, this gives
\begin{equation}
\mathcal{K}_{xy}^{GG}[\phi,\Delta]\ =\ \mathcal{K}^{GG}_{xy}[\varphi,\mathcal{G}]\:-\:2\,\hbar\,\frac{\delta \Gamma^{(\mathrm{HF})}_2[\phi,\Delta]}{\delta \Delta_{xy}^{GG}}\bigg|_{\varphi,\mathcal{G}}\ =\ -\:2\,\hbar\,\frac{\delta \Gamma^{(\mathrm{HF})}_2[\phi,\Delta]}{\delta\Delta^{HH}_{xy}}\bigg|_{\varphi,\mathcal{G}}\;,
\end{equation}
from which eq.~\eqref{eq:GGsource} immediately follows.
Note that the right-hand side of eq.~\eqref{eq:GGsource} does vanish in the symmetric phase, and the standard Schwinger-Dyson equation again holds.

In the SSB phase, we may expand the background field configuration of the Higgs as $\varphi_x^H=v+\hbar\,\delta\varphi^H_x$, where $v=\pm\,|m|/\sqrt{\lambda}$ is the tree-level vacuum expectation value and
\begin{equation}
\label{eq:tadH}
\delta \varphi^H_x\ =\ -\:\lambda\,v\, \mathcal{G}_{xy}^{HH}\Big(3\,\mathcal{G}_{yy}^{HH}+\mathcal{G}_{yy}^{GG}\Big)\ =\ \frac{\lambda\, v}{2m^2}\,\Big(3\,\mathcal{G}_{xx}^{HH}+\mathcal{G}_{xx}^{GG}\Big)\:+\:\mathcal{O}(\hbar)\;.
\end{equation}
By applying the identity in eq.~\eqref{eq:Ktadconst}, the $\hbar$ corrections to the Goldstone-boson inverse two-point function $\mathcal{G}^{-1,\,GG}_{xy}$ in eq.~\eqref{eq:GGG} are given by
\begin{equation}
\label{eq:GGGhbar}
\mathcal{G}^{-1,\,GG}_{xy}\ \supset\ 2\,\hbar\,\lambda\, v\, \delta\varphi^H_x\,\delta^{(4)}_{xy}\:+\:2\,\hbar\,\frac{\delta \Gamma^{(\mathrm{HF})}_2[\phi,\Delta]}{\delta \Delta^{HH}_{xy}}\bigg|_{\varphi,\mathcal{G}}\;.
\end{equation}
We emphasize that the tadpole self-energy appearing in eq.~\eqref{eq:GGGhbar} is that of the Higgs boson, by virtue of eq.~\eqref{eq:Ktadconst}, \emph{not} the Goldstone boson. The tadpole self-energy of the Higgs is given by
\begin{equation}
\label{eq:HFH}
2\,\frac{\delta \Gamma_2^{({\rm HF})}[\phi,\Delta]}{\delta \Delta^{HH}_{xy}}\bigg|_{\varphi,\mathcal{G}}\ =\ \lambda\, \Big(3\,\mathcal{G}_{xx}^{HH}\:+\:\mathcal{G}_{xx}^{GG}\Big)\delta^{(4)}_{xy}\;.
\end{equation}
For comparison, the corresponding would-be self-energy of the Goldstone boson is given by
\begin{equation}
2\,\frac{\delta \Gamma_2^{({\rm HF})}[\phi,\Delta]}{\delta \Delta^{GG}_{xy}}\bigg|_{\varphi,\mathcal{G}}\ =\ \lambda\, \Big(\mathcal{G}_{xx}^{HH}\:+\:3\,\mathcal{G}_{xx}^{GG}\Big)\delta^{(4)}_{xy}\;,
\end{equation}
in which the combinatorical factors are interchanged relative to eq.~\eqref{eq:HFH}.
Substituting eqs.~\eqref{eq:tadH} and~\eqref{eq:HFH} into eq.~\eqref{eq:GGGhbar} and using the fact that $\lambda v^2=-\,m^2$, we find that the order $\hbar$ corrections exactly cancel. Therefore, we arrive at the result
\begin{equation}
\mathcal{G}^{-1,\,GG}_{xy}\ =\ G^{-1,\,GG}_{xy}\ =\ -\:\delta^{(4)}_{xy}\partial_x^2\;,
\end{equation}
corresponding to a \emph{massless} Goldstone boson. Had we instead imposed the on-shell condition for the Goldstone mode, we would have found
\begin{equation}
\mathcal{G}^{-1,\,GG}_{xy}\ =\  \delta^{(4)}_{xy}\Big[-\,\partial_x^2\:-\:2\,\hbar\,\lambda\,\Big(\mathcal{G}^{HH}_{xx}\:-\:\mathcal{G}^{GG}_{xx}\Big)\Big]\;,
\end{equation}
in which the pathological Goldstone-boson mass has arisen. It is interesting to remark that this algebraic cancellation of the $\hbar$ corrections to the Goldstone mass bears resemblance to the cancellation that occurs in OPT methods (cf.~ref.~\cite{Duarte:2011ph}). It is well known that this cancellation of the mass contributions to the Goldstone boson occurs in the full one-loop calculation at $T=0$. Nonetheless, despite having used the HF approximation, the result in eq.~\eqref{eq:tadH} is correct to order $\lambda$ as the truncation has no effect on the form of $\Gamma_1[\phi,\Delta]$. Together with the preservation of Goldstone's theorem, this method of external sources thus recovers the salient features of this simple model of SSB in the HF approximation.

We will now consider the thermal mass corrections to the Higgs-boson inverse two-point function in order to illustrate that we also obtain the correct second-order phase transition at finite temperature within this modified HF truncation. We remark here that information about the statistical ensemble may be encoded in the external sources in the usual way (see e.g.~refs.~\cite{Calzetta:1986cq,Berges:2004yj,Millington:2012pf}) without affecting the preceding discussions, since the statistical parts amount to additive corrections to the external sources.

The Higgs-boson mass corrections are given by
\begin{equation}
\label{eq:GHHhbar}
\mathcal{G}^{-1,\,HH}_{xy}\ \supset \ 6\,\hbar\,\lambda\,v\,\delta\varphi_x^H\:+\:2\,\hbar\,\frac{\delta \Gamma^{({\rm HF})}_2[\phi,\Delta]}{\delta \Delta^{HH}_{xy}}\bigg|_{\varphi,\mathcal{G}}\ = \ -\:2\,\hbar\,\lambda\,\Big(3\,\mathcal{G}_{xx}^{HH}+\mathcal{G}_{xx}^{GG}\Big)\delta^{(4)}_{xy}\;.
\end{equation}
Working at finite temperature, taking $T\gg |m|$, the thermal contributions to the tadpole corrections are obtained using
\begin{equation}
\mathcal{G}^{HH}_{xx}\big|_{\mathrm{therm}}\ \approx \ \mathcal{G}^{GG}_{xx}\big|_{\mathrm{therm}}\ \approx\ \int\!\!\frac{\mathrm{d}^3\mathbf{k}}{(2\pi)^3}\,\frac{1}{|\mathbf{k}|}\,\frac{1}{e^{|\mathbf{k}|/T}-1}\ =\ \frac{T^2}{12}\;.
\end{equation}
Substituting this into eq.~\eqref{eq:GHHhbar}, we find the thermal mass of the Higgs boson in the Hartree-Fock approximation ($\hbar=1$)
\begin{equation}
m_H^2\ =\ -\:2m^2\:-\:\frac{8\,\lambda \,T^2}{12}\;,
\end{equation}
which is in agreement with the result presented in ref.~\cite{Pilaftsis:2013xna}. The critical temperature is obtained by setting both the Higgs and Goldstone masses to zero, and we obtain
\begin{equation}
T_c\ =\ \sqrt{3}\,|v|\;,
\end{equation}
where we emphasize that $v$ is the tree-level vacuum expectation value.
The mass gap equations of the Higgs and Goldstone modes are given the Schwinger-Dyson equations in the limit of vanishing momenta:
\begin{subequations}
\label{eq:massgap}
\begin{gather}
m_H^2\ =\ 3\,\lambda\,(v_{\rm HF})^2\:+\:m^2\:+\:\lambda\,\Big(3\,\mathcal{G}^{HH}_{xx}\:+\:\mathcal{G}^{GG}_{xx}\Big)\;,\\
m_G^2\ =\ \lambda\,(v_{\rm HF})^2\:+\:m^2\:+\:\lambda\,\Big(3\,\mathcal{G}^{HH}_{xx}\:+\:\mathcal{G}^{GG}_{xx}\Big)\;,
\end{gather}
\end{subequations}
where $v_{\rm HF}$ is the one-loop vacuum expectation value in the HF approximation and it is understood that the dressed propagators $\mathcal{G}_{xx}^{HH}$ and $\mathcal{G}_{xx}^{GG}$ are functions of the one-loop masses $m_H$ and $m_G$. We emphasize that, by virtue of the constraint on the bi-local sources in eq.~\eqref{eq:Ktadconst}, the Higgs tadpole self-energy appears in the gap equations for \emph{both} the Higgs and Goldstone modes and, as a result, it immediately follows from eq.~\eqref{eq:massgap} that
\begin{equation}
v_{\rm HF}^2\ =\ \frac{m_H^2-m_G^2}{2\lambda}\;.
\end{equation}
Since $m_H^2=m_G^2=0$ at the critical temperature, we simultaneously have $v^2_{\rm HF}=0$ and therefore find a second-order phase transition in agreement with ref.~\cite{Pilaftsis:2013xna}. We may verify this explicitly for the case in which $T_c\gg |m|$, since
\begin{equation}
m^2\:+\:\lambda\,\Big(3\,\mathcal{G}^{HH}_{xx}\:+\:\mathcal{G}^{GG}_{xx}\Big)\Big|_{T=T_c}\ =\ m^2\:+\:\frac{\lambda\,T_c^2}{3}\ =\ 0\;.
\end{equation}
Herein, the $T=0$ parts are irrelevant, since they are zero at the critical temperature by virtue of the fact that $m_H^2=m_G^2=0$ (cf.~ref.~\cite{Pilaftsis:2013xna}). Before concluding this section, we remark more explicitly on the methodological difference between the present approach and that of ref.~\cite{Pilaftsis:2013xna}. This comparison is most easily made by considering the mass gap equations in eq.~\eqref{eq:massgap}. Due to the common combinatorics appearing with the loop corrections, which result from the constraint in eq.~\eqref{eq:Ktadconst}, the mass gap equations can be solved analytically to show that the phase transition is second order. On the other hand, in the approach of ref.~\cite{Pilaftsis:2013xna}, the combinatorics in the Goldstone mass gap equation are reversed compared to the Higgs mass gap equation, such that the system, along with the additional constraint from the Ward identities, must be solved self-consistently. The results of both analyses are, however, entirely equivalent.

The discussions of this section are intended to illustrate the potential utility of this alternative method of evaluating the effective action and are not intended to constitute a complete treatment of global-symmetry preservation in truncations of the effective action. It is for this reason that we have not discussed the threshold structure of the Higgs and Goldstone self-energies beyond the HF approximation, although it is expected that these will behave correctly, since the Goldstone bosons appearing in the loops are massless as in the PT symmetry-improved effective action~\cite{Pilaftsis:2013xna} and in contrast to the external propagator method~\cite{vanHees:2002bv}. Nevertheless, by using this method of external sources, we have been able to ensure that the Goldstone boson remains massless in the HF approximation and obtain the correct second-order phase transition. With this success in mind, it would be of interest to consider the behaviour of IR divergences in this construction, which have been shown to be absent in the symmetry-improved effective action (see~ref.~\cite{Pilaftsis:2015cka,Pilaftsis:2015bbs}).  We also remark that it would be interesting to consider higher-orders in the $1/N$ expansion~\cite{'tHooft:1973jz}, where it is known that the 2PPI effective action yields massive Goldstone bosons at the next order in $1/N$~\cite{Baacke:2002pi}. By combining the external-source method as applied to the 2PPI effective in section~\ref{sec:VC} with the use of symmetry constraints highlighted in this section, it is anticipated that this situation may also be rectified in complete analogy to the example of the HF approximation above.

\section{Conclusions}
\label{sec:conc}

We have described an alternative method of evaluating the effective action, where, in contrast to the usual approach, the physical limit is obtained in the presence of \emph{non-vanishing} external sources in vacuum. These external sources may then be used to constrain the effective action.

We have illustrated the utility of this general approach by means of three concrete examples:
\begin{enumerate}
\item[(i)] By forcing the system to follow its extremal quantum trajectory, we are able to recover the CJT effective action. This approach, however, has the advantage that the saddle-point evaluation of the effective action is performed along the quantum path, thereby being of relevance to problems in which the quantum and classical paths are non-perturbatively far away from each other. This methodology is of much significance to studies of false vacuum decay, as was emphasised in the complementary work presented in ref.~\cite{Garbrecht:2015yza}, and it may serve as a theoretical foundation for metastability calculations in the electroweak sector of the SM, where radiative corrections have a pivotal impact upon the vacuum structure.

\item[(ii)] We have demonstrated how the external sources may be used to recover variants of the effective action in the particular case of the CV 2PPI effective action, with the advantage that we do not need to isolate terms in the effective action artificially in order to avoid problems of double counting.

\item[(iii)] We have illustrated that this approach may be used to constrain truncations of the effective action so as to preserve symmetry properties. In particular, we have described how this method may be used to re-organize the HF approximation of a globally $O(2)$ invariant model with SSB, such that the Goldstone boson remains massless algebraically and we obtain the correct second-order phase transition.

\end{enumerate}

Aside from the additional studies in the context of global symmetries highlighted in section~\ref{sec:syms}, it is also of interest to generalize this approach to the case of Abelian and non-Abelian gauge theories. For instance, one might ensure that the photon remains massless in truncations of the 2PI effective action of QED by constraining the longitudinal component of the bi-local gauge source by the Ward-Takahashi identities or, analogously, obtain massless gluons in QCD by using the Slavnov-Taylor identities to constrain the bi-local gauge source. In this way, one might remove the pathological gauge dependency that results in truncations of the effective action.\footnote{This idea has been pursued recently by the authors of ref.~\cite{Plascencia:2015pga} in the context of tunneling rates, where it is argued that evaluation of the tunneling path integral along the quantum path directly will ensure gauge independence of the tunneling rate.}

\section*{Acknowledgments}
P.M. would like to thank Apostolos Pilaftsis and Daniele Teresi for many illuminating discussions of effective-action techniques both at zero and finite temperature and in the context of global symmetries. The authors would like to thank Paul Saffin, Carlos Tamarit and Daniele Teresi for constructive comments and discussions. The work  of P.M.~was supported in part by the Science and Technologies Facilities Council (STFC) under grant no.~ST/L000393/1 and a University  Foundation Fellowship
(TUFF) from  the Technische  Universit\"{a}t M\"{u}nchen. The  work of
B.G.~is  supported by the  Gottfried Wilhelm Leibniz Programme  of the
Deutsche  Forschungsgemeinschaft  (DFG).    Both  authors  acknowledge
support from the DFG cluster of excellence Origin and Structure of the
Universe.


\begin{thebibliography}{99}

\bibitem{Jackiw:1974cv}
  R.~Jackiw,
  Phys.\ Rev.\ D {\bf 9} (1974) 1686.

\bibitem{Cornwall:1974vz}
  J.~M.~Cornwall, R.~Jackiw and E.~Tomboulis,
  Phys.\ Rev.\ D {\bf 10} (1974) 2428.
  
\bibitem{Arrizabalaga:2002hn}
  A.~Arrizabalaga and J.~Smit,
  Phys.\ Rev.\ D {\bf 66} (2002) 065014
  [hep-ph/0207044].
  
\bibitem{Carrington:2003ut}
  M.~E.~Carrington, G.~Kunstatter and H.~Zaraket,
  Eur.\ Phys.\ J.\ C {\bf 42} (2005) 253
  [hep-ph/0309084].

\bibitem{Mottola:2003vx}
  E.~Mottola,
  in \emph{Strong and electroweak matter 2002: proceedings of the SEWM 2002 meeting}. World Scientific, Singapore (2003). 
  [hep-ph/0304279].

\bibitem{Ward:1950xp}
  J.~C.~Ward,
  Phys.\ Rev.\  {\bf 78} (1950) 182.
  
\bibitem{Takahashi:1957xn}
  Y.~Takahashi,
  Nuovo Cim.\  {\bf 6} (1957) 371.

\bibitem{Taylor:1971ff}
  J.~C.~Taylor,
  Nucl.\ Phys.\ B {\bf 33} (1971) 436.
  
\bibitem{Slavnov:1972fg}
  A.~A.~Slavnov,
  Theor.\ Math.\ Phys.\  {\bf 10} (1972) 99
   [Teor.\ Mat.\ Fiz.\  {\bf 10} (1972) 153].
  
\bibitem{Goldstone:1961eq}
  J.~Goldstone,
  Nuovo Cim.\  {\bf 19} (1961) 154.
  
\bibitem{Goldstone:1962es}
  J.~Goldstone, A.~Salam and S.~Weinberg,
  Phys.\ Rev.\  {\bf 127} (1962) 965.

\bibitem{Baym:1977qb}
  G.~Baym and G.~Grinstein,
  Phys.\ Rev.\ D {\bf 15} (1977) 2897.
  
\bibitem{AmelinoCamelia:1997dd}
  G.~Amelino-Camelia,
  Phys.\ Lett.\ B {\bf 407} (1997) 268
  [hep-ph/9702403].
  
\bibitem{Petropoulos:1998gt}
  N.~Petropoulos,
  J.\ Phys.\ G {\bf 25} (1999) 2225
  [hep-ph/9807331].
  
\bibitem{Lenaghan:1999si}
  J.~T.~Lenaghan and D.~H.~Rischke,
  J.\ Phys.\ G {\bf 26} (2000) 431
  [nucl-th/9901049].

\bibitem{Baacke:2002pi}
  J.~Baacke and S.~Michalski,
  Phys.\ Rev.\ D {\bf 67} (2003) 085006
  [hep-ph/0210060].

\bibitem{Ivanov:2005yj}
  Y.~B.~Ivanov, F.~Riek and J.~Knoll,
  Phys.\ Rev.\ D {\bf 71} (2005) 105016
  [hep-ph/0502146].
  
\bibitem{Ivanov:2005bv}
  Y.~B.~Ivanov, F.~Riek, H.~van Hees and J.~Knoll,
  Phys.\ Rev.\ D {\bf 72} (2005) 036008
  [hep-ph/0506157].
  
\bibitem{Seel:2011ju}
  E.~Seel, S.~Struber, F.~Giacosa and D.~H.~Rischke,
  Phys.\ Rev.\ D {\bf 86} (2012) 125010
  [arXiv:1108.1918 [hep-ph]].
  
\bibitem{Grahl:2011yk}
  M.~Grahl, E.~Seel, F.~Giacosa and D.~H.~Rischke,
  Phys.\ Rev.\ D {\bf 87} (2013) 9,  096014
  [arXiv:1110.2698 [nucl-th]].
  
\bibitem{Marko:2013lxa}
  G.~Mark\'{o}, U.~Reinosa and Z.~Sz\'{e}p,
  Phys.\ Rev.\ D {\bf 87} (2013) 10,  105001
  [arXiv:1303.0230 [hep-ph]].
  
\bibitem{Nemoto:1999qf}
  Y.~Nemoto, K.~Naito and M.~Oka,
  Eur.\ Phys.\ J.\ A {\bf 9} (2000) 245
  [hep-ph/9911431].

\bibitem{vanHees:2002bv}
  H.~van Hees and J.~Knoll,
  Phys.\ Rev.\ D {\bf 66} (2002) 025028
  [hep-ph/0203008].
  
\bibitem{Chiku:1998kd}
  S.~Chiku and T.~Hatsuda,
  Phys.\ Rev.\ D {\bf 58} (1998) 076001
  [hep-ph/9803226].  
  
\bibitem{Duarte:2011ph}
  D.~C.~Duarte, R.~L.~S.~Farias and R.~O.~Ramos,
  Phys.\ Rev.\ D {\bf 84} (2011) 083525
  [arXiv:1108.4428 [hep-ph]].
  
\bibitem{Pilaftsis:2013xna}
  A.~Pilaftsis and D.~Teresi,
  Nucl.\ Phys.\ B {\bf 874} (2013) 2,  594
  [arXiv:1305.3221 [hep-ph]].

\bibitem{Tetradis:1992xd}
  N.~Tetradis and C.~Wetterich,
  Nucl.\ Phys.\ B {\bf 398} (1993) 659.
  
\bibitem{Pilaftsis:2015cka}
  A.~Pilaftsis and D.~Teresi,
  J.\ Phys.\ Conf.\ Ser.\  {\bf 631} (2015) 1,  012008
  [arXiv:1502.07986 [hep-ph]].
  
\bibitem{Pilaftsis:2015bbs}
  A.~Pilaftsis and D.~Teresi,
  arXiv:1511.05347 [hep-ph].
  
\bibitem{Reinosa:2007vi}
  U.~Reinosa and J.~Serreau,
  JHEP {\bf 0711} (2007) 097
  [arXiv:0708.0971 [hep-th]].
  
\bibitem{Schwinger:1961} 
  J.~S.~Schwinger,
  J.\ Math.\ Phys.\  {\bf 2}, 407 (1961).

\bibitem{Keldysh:1964} 
  L.~V.~Keldysh,
  Zh.\ Eksp.\ Teor.\ Fiz.\  {\bf 47}, 1515 (1964)
  [Sov.\ Phys.\ JETP {\bf 20}, 1018 (1965)].  

\bibitem{Norton:1974bm}
  R.~E.~Norton and J.~M.~Cornwall,
  Annals Phys.\  {\bf 91} (1975) 106.

\bibitem{Jordan:1986ug}
  R.~D.~Jordan,
  Phys.\ Rev.\ D {\bf 33} (1986) 444.

\bibitem{Calzetta:1986ey}
   E.~Calzetta and B.~L.~Hu,
   Phys.~Rev.~D {\bf 35}, 495 (1987).

\bibitem{Calzetta:1986cq}
  E.~Calzetta and B.~L.~Hu,
  Phys.~Rev.~D {\bf 37}, 2878 (1988).
  
\bibitem{AmelinoCamelia:1992nc}
  G.~Amelino-Camelia and S.~Y.~Pi,
  Phys.\ Rev.\ D {\bf 47} (1993) 2356
  [hep-ph/9211211].
  
\bibitem{Baym:1961zz}
  G.~Baym and L.~P.~Kadanoff,
  Phys.\ Rev.\  {\bf 124} (1961) 287.

\bibitem{KB} 
  L. Kadanoff and G. Baym,
  {\it Quantum Statistical Mechanics}, Benjamin, New York (1962).
  
\bibitem{Blaizot:2001nr}
  J.~P.~Blaizot and E.~Iancu,
  Phys.\ Rept.\  {\bf 359} (2002) 355
  [hep-ph/0101103].
  
\bibitem{Prokopec:2003pj}
  T.~Prokopec, M.~G.~Schmidt and S.~Weinstock,
  Annals Phys.\  {\bf 314} (2004) 208
  [hep-ph/0312110].

\bibitem{Prokopec:2004ic}
  T.~Prokopec, M.~G.~Schmidt and S.~Weinstock,
  Annals Phys.\  {\bf 314} (2004) 267
  [hep-ph/0406140].

\bibitem{Berges:2004yj}
  J.~Berges,
  AIP Conf.\ Proc.\  {\bf 739} (2005) 3
  [hep-ph/0409233].
  
\bibitem{Millington:2012pf}
  P.~Millington and A.~Pilaftsis,
  Phys.\ Rev.\ D {\bf 88} (2013) 8,  085009
  [arXiv:1211.3152 [hep-ph]].
  
\bibitem{vanHees:2001ik}
  H.~van Hees and J.~Knoll,
  Phys.\ Rev.\ D {\bf 65} (2002) 025010
  [hep-ph/0107200].
  
\bibitem{VanHees:2001pf}
  H.~Van Hees and J.~Knoll,
  Phys.\ Rev.\ D {\bf 65} (2002) 105005
  [hep-ph/0111193].


\bibitem{Blaizot:2003br}
  J.~P.~Blaizot, E.~Iancu and U.~Reinosa,
  Phys.\ Lett.\ B {\bf 568} (2003) 160
  [hep-ph/0301201].

\bibitem{Blaizot:2003an}
  J.~P.~Blaizot, E.~Iancu and U.~Reinosa,
  Nucl.\ Phys.\ A {\bf 736} (2004) 149
  [hep-ph/0312085].
  
\bibitem{Berges:2004hn}
  J.~Berges, S.~Borsanyi, U.~Reinosa and J.~Serreau,
  Phys.\ Rev.\ D {\bf 71} (2005) 105004
  [hep-ph/0409123].

\bibitem{Berges:2005hc}
  J.~Berges, S.~Borsanyi, U.~Reinosa and J.~Serreau,
  Annals Phys.\  {\bf 320} (2005) 344
  [hep-ph/0503240].
  
\bibitem{Aad:2012tfa}
  G.~Aad {\it et al.}  [ATLAS Collaboration],
  Phys.\ Lett.\ B {\bf 716} (2012) 1
  [arXiv:1207.7214 [hep-ex]].

\bibitem{Chatrchyan:2012ufa}
  S.~Chatrchyan {\it et al.}  [CMS Collaboration],
  Phys.\ Lett.\ B {\bf 716} (2012) 30
  [arXiv:1207.7235 [hep-ex]].
  
\bibitem{Agashe:2014kda}
  K.~A.~Olive {\it et al.} [Particle Data Group Collaboration],
  Chin.\ Phys.\ C {\bf 38} (2014) 090001.
  
\bibitem{Cabibbo:1979ay}
  N.~Cabibbo, L.~Maiani, G.~Parisi and R.~Petronzio,
  Nucl.\ Phys.\ B {\bf 158} (1979) 295.

\bibitem{Sher:1988mj}
  M.~Sher,
  Phys.\ Rept.\  {\bf 179} (1989) 273.

\bibitem{Sher:1993mf}
  M.~Sher,
  Phys.\ Lett.\ B {\bf 317} (1993) 159
   [Addendum-ibid.\ B {\bf 331} (1994) 448]
  [hep-ph/9307342].

\bibitem{Isidori:2001bm}
  G.~Isidori, G.~Ridolfi and A.~Strumia,
  Nucl.\ Phys.\ B {\bf 609} (2001) 387
  [hep-ph/0104016].

\bibitem{EliasMiro:2011aa}
  J.~Elias-Mir\'{o}, J.~R.~Espinosa, G.~F.~Giudice, G.~Isidori,
  A.~Riotto and A.~Strumia,
  Phys.\ Lett.\ B {\bf 709} (2012) 222
  [arXiv:1112.3022 [hep-ph]].

\bibitem{Degrassi:2012ry}
  G.~Degrassi, S.~Di Vita, J.~Elias-Mir\'{o}, J.~R.~Espinosa, G.~F.~Giudice,
  G.~Isidori and A.~Strumia,
  JHEP {\bf 1208} (2012) 098
  [arXiv:1205.6497 [hep-ph]].

\bibitem{Alekhin:2012py}
  S.~Alekhin, A.~Djouadi and S.~Moch,
  Phys.\ Lett.\ B {\bf 716} (2012) 214
  [arXiv:1207.0980 [hep-ph]].

\bibitem{Buttazzo:2013uya}
  D.~Buttazzo, G.~Degrassi, P.~P.~Giardino, G.~F.~Giudice, F.~Sala,
  A.~Salvio and A.~Strumia,
  JHEP {\bf 1312} (2013) 089
  [arXiv:1307.3536 [hep-ph]].
  
\bibitem{Bednyakov:2015sca}
  A.~V.~Bednyakov, B.~A.~Kniehl, A.~F.~Pikelner and O.~L.~Veretin,
  arXiv:1507.08833 [hep-ph].
  
\bibitem{DiLuzio:2015iua}
  L.~Di Luzio, G.~Isidori and G.~Ridolfi,
  Phys.\ Lett.\ B {\bf 753} (2016) 150
  [arXiv:1509.05028 [hep-ph]].
  
\bibitem{Bezrukov:2012sa}
  F.~Bezrukov, M.~Y.~Kalmykov, B.~A.~Kniehl and M.~Shaposhnikov,
  JHEP {\bf 1210} (2012) 140
  [arXiv:1205.2893 [hep-ph]].

\bibitem{Masina:2012tz}
  I.~Masina,
  Phys.\ Rev.\ D {\bf 87} (2013) 053001
  [arXiv:1209.0393 [hep-ph]].
  
\bibitem{Branchina:2013jra}
  V.~Branchina and E.~Messina,
  Phys.\ Rev.\ Lett.\  {\bf 111} (2013) 241801
  [arXiv:1307.5193 [hep-ph]].

\bibitem{Branchina:2014usa}
  V.~Branchina, E.~Messina and A.~Platania,
  JHEP {\bf 1409} (2014) 182
  [arXiv:1407.4112 [hep-ph]].

\bibitem{Branchina:2014rva}
  V.~Branchina, E.~Messina and M.~Sher,
  Phys.\ Rev.\ D {\bf 91} (2015) 013003
  [arXiv:1408.5302 [hep-ph]].
  
\bibitem{Lalak}
  Z.~Lalak, M.~Lewicki and P.~Olszewski,
  JHEP {\bf 1405} (2014) 119
  [arXiv:1402.3826 [hep-ph]].
  
\bibitem{Eichhorn:2015kea}
  A.~Eichhorn, H.~Gies, J.~Jaeckel, T.~Plehn, M.~M.~Scherer and R.~Sondenheimer,
  [arXiv:1501.02812 [hep-ph]].

\bibitem{Branchina:2015nda}
  V.~Branchina and E.~Messina,
  arXiv:1507.08812 [hep-ph].
  
\bibitem{Gies:2014xha}
  H.~Gies and R.~Sondenheimer,
  Eur.\ Phys.\ J.\ C {\bf 75} (2015) 2,  68
  [arXiv:1407.8124 [hep-ph]].
  
\bibitem{Weinberg:1992ds}
  E.~J.~Weinberg,
  Phys.\ Rev.\ D {\bf 47} (1993) 4614
  [hep-ph/9211314].

\bibitem{Coleman:1973jx}
  S.~R.~Coleman and E.~J.~Weinberg,
  Phys.\ Rev.\ D {\bf 7} (1973) 1888.

\bibitem{Kirzhnits:1972ut}
  D.~A.~Kirzhnits and A.~D.~Linde,
  Phys.\ Lett.\ B {\bf 42} (1972) 471.

\bibitem{Dolan:1973qd}
  L.~Dolan and R.~Jackiw,
  Phys.\ Rev.\ D {\bf 9} (1974) 3320.

\bibitem{Weinberg:1974hy}
  S.~Weinberg,
  Phys.\ Rev.\ D {\bf 9} (1974) 3357.  

\bibitem{Verschelde:1992bs}
  H.~Verschelde and M.~Coppens,
  Phys.\ Lett.\ B {\bf 287} (1992) 133.
  
\bibitem{Verschelde:1992ig}
  H.~Verschelde and M.~Coppens,
  Phys.\ Lett.\ B {\bf 295} (1992) 83.
 
\bibitem{Coppens:1993zc}
  M.~Coppens and H.~Verschelde,
  Z.\ Phys.\ C {\bf 58} (1993) 319.
  
\bibitem{Coppens:1993ri}
  M.~Coppens and H.~Verschelde,
  Z.\ Phys.\ C {\bf 57} (1993) 349.
  
\bibitem{Hartree}
  D.~Hartree,
  Math.~Proc.~Cam.~Phil.~Soc.~24 (1928) 89--132.

\bibitem{Fock}
  V.~Fock, Z.~Physik 61 (1930) 126--148.

\bibitem{Carrington:2004sn}
  M.~E.~Carrington,
  Eur.\ Phys.\ J.\ C {\bf 35} (2004) 383
  [hep-ph/0401123].
  
\bibitem{Garbrecht:2015oea}
  B.~Garbrecht and P.~Millington,
  Phys.\ Rev.\ D {\bf 91} (2015) 105021
  [arXiv:1501.07466 [hep-th]].
  
\bibitem{Coleman:1977py}
  S.~R.~Coleman,
  Phys.\ Rev.\ D {\bf 15} (1977) 2929
   [Erratum-ibid.\ D {\bf 16} (1977) 1248].

\bibitem{Callan:1977pt}
  C.~G.~Callan, Jr. and S.~R.~Coleman,
  Phys.\ Rev.\ D {\bf 16} (1977) 1762.
  
\bibitem{Weinberg:1987vp}
  E.~J.~Weinberg and A.~Wu,
  Phys.\ Rev.\ D {\bf 36} (1987) 2474.
  
\bibitem{Garbrecht:2015yza}
  B.~Garbrecht and P.~Millington,
  Phys.\ Rev.\ D {\bf 92} (2015) 12,  125022
  [arXiv:1509.08480 [hep-ph]].
  
\bibitem{Englert:1964et}
  F.~Englert and R.~Brout,
  Phys.\ Rev.\ Lett.\  {\bf 13} (1964) 321.

\bibitem{Higgs:1964pj}
  P.~W.~Higgs,
  Phys.\ Rev.\ Lett.\  {\bf 13} (1964) 508.

\bibitem{Guralnik:1964eu}
  G.~S.~Guralnik, C.~R.~Hagen and T.~W.~B.~Kibble,
  Phys.\ Rev.\ Lett.\  {\bf 13} (1964) 585.
  
\bibitem{Weinberg:1971fb}
  S.~Weinberg,
  Phys.\ Rev.\ Lett.\  {\bf 27} (1971) 1688.

\bibitem{Weinberg:1973ew}
  S.~Weinberg,
  Phys.\ Rev.\ D {\bf 7} (1973) 1068.
  
\bibitem{'tHooft:1973jz}
  G.~'t Hooft,
  Nucl.\ Phys.\ B {\bf 72} (1974) 461.
  
\bibitem{Plascencia:2015pga}
  A.~D.~Plascencia and C.~Tamarit,
  arXiv:1510.07613 [hep-ph].

\end{thebibliography}
\end{document}